\documentclass[journal]{IEEEtran}
\usepackage{graphicx,natbib,xcolor,colortbl} 
\usepackage{yhmath}
\usepackage{algorithmicx,algorithm,algpseudocode}
\usepackage{natbib}
\usepackage[caption=false]{subfig} 
\usepackage{amsmath}
\usepackage{multirow}
\usepackage{booktabs}
\usepackage[flushleft]{threeparttable} 
\usepackage{subeqnarray}
\usepackage{cases}
\usepackage{amsthm} 
\usepackage[cmex10]{mathtools}
\usepackage{cases}
\usepackage[figuresright]{rotating}
\usepackage{amsfonts,amssymb}

\begin{document}

\title{A heuristic method for data allocation and task scheduling on heterogeneous multiprocessor systems under memory constraints}


\author{Junwen Ding,
        Liangcai Song,
        Siyuan Li,
        Chen Wu,
        Ronghua He,
        Zhouxing Su,
        and Zhipeng L\"u*,
\thanks{J. Ding, L. Song, S. Li, Z. Su, and Z. L\"u are with Huazhong University of Science and Technology, Wuhan, China, e-mail: \{junwending,zhipeng.lv\}@hust.edu.cn (Corresponding author: Zhipeng L\"u.)}
\thanks{C. Wu and R. He are with 2012 Lab, Huawei Technologies Co., Ltd. Shenzhen, China.}
}

\markboth{}
{Ding \MakeLowercase{\textit{et al.}}: Bare Demo of IEEEtran.cls for IEEE Journals}

\maketitle

\begin{abstract}
Computing workflows in heterogeneous multiprocessor systems are frequently modeled as directed acyclic graphs of tasks and data blocks, which represent computational modules and their dependencies in the form of data produced by a task and used by others. However, for some workflows, such as the task schedule in a digital signal processor may run out of memory by exposing too much parallelism. This paper focuses on the data allocation and task scheduling problem under memory constraints, and concentrates on shared memory platforms. We first propose an integer linear programming model to formulate the problem. Then we consider the problem as an extended flexible job shop scheduling problem, while trying to minimize the critical path of the graph. To solve this problem, we propose a tabu search algorithm (TS) which combines several distinguished features such as a greedy initial solution construction method and a mixed neighborhood evaluation strategy based on exact evaluation and approximate evaluation methods. Experimental results on randomly generated instances show that the the proposed TS algorithm can obtain relatively high-quality solutions in a reasonable computational time. In specific, the tabu search method averagely improves the makespan by 5-25\% compared to the classical load balancing algorithm that are widely used in the literature. Besides, some key features of TS are also analyzed to identify its success factors.
\end{abstract}


\IEEEpeerreviewmaketitle

\section{Introduction}

\IEEEPARstart{A} digital signal processor (DSP) is a specialized microprocessor chip with delicately refined architecture for the operational requirements of digital signal processing \citep{chantem2010temperature}. DSPs are fabricated on metal oxide semiconductor (MOS) integrated circuit chips. They are widely used in audio signal processing, digital image processing, speech recognition systems, high performance computing centers, and in common consumer electronic devices such as mobile phones, notebook computers, smart watches, and intelligent Wearable device \citep{baruah2006partitioned}.

There are different types of cores and memories on DSP chip, where the core is the unit that performs calculations and memory is the unit that stores data. Similar to the description in \cite{chen2012online}, core types include general purpose core and synergistic processor core, and memory types include high-speed memory and low-speed memory such as DDR. The cores are organized according to the cluster and group levels, where each group corresponds to a local high-speed memory, while other high-speed memory and low-speed memory are shared globally.

Parallel computing tasks on multiprocessor systems are often modeled by Directed Acyclic task Graphs (DAG), where nodes and edges respectively represent tasks and the dependencies between tasks \citep{chiang2006multi,du2013efficient}. Given a series of tasks to be executed on a DSP processor, and the data blocks generated by tasks, i.e., the dependencies between tasks, the task scheduling problem is to assign each task to the cores, specify the storage location for the data block, and also determine the execution order of the tasks on each core, where the objective is to minimize the total completion time of all tasks and the usage of high-speed memory, and improve the utilization of the cores and memories.

The job shop scheduling problem is a fundamental problem in the fields of intelligent manufacturing and high-performance computing, which mainly studies how to schedule priority resources to execute multiple tasks in sequence, so that the maximum completion time of all tasks is minimized. For example,
when a chip foundry produces chips, each wafer undergoes multiple processes such as photolithography and etching on different machines sequentially \citep{yin2018tasks}.
In some large-scale parallel computing scenarios, there are dependencies between computing tasks, and the input of the successor task is the output of the predecessor task \citep{ilavarasan2007low}.

The scheduling problems in the actual production process are often more complex, since there are various constraints from different dimensions need to be considered apart from scheduling computing resources. For example,
when multi-core processors in a cloud computing data center are shared by a large number of parallel tasks, it is necessary to allocate cores to tasks and schedule tasks simultaneously under energy constraints or performance constraints \citep{kang2011task}.
In parallel computing scenarios, in addition to the occupancy of computing resources, it is also necessary to consider that the memory resources occupied by concurrent computing tasks cannot exceed the maximum capacity limit.
Heterogeneous chips integrate different computing units to allow each computing unit to perform compatible tasks, which arises higher requirements for task scheduling since there are more complex constraints among different types of memories \citep{kang2010darts}.

As heterogeneous processors is prevalent due to its high efficiency, the same type of operations can be processed by different cores with different processing time and data capacity. Furthermore, different components of a distributed shared-memory show significant heterogeneity in data access time \citep{lakshmanan2009coordinated,ouni2011partitioning,wang2014energy}. Therefore, several important issues are arisen and need to be resolved, i.e., how to assign each computational task to a proper processor; how to allocate each datum to a proper memory; and how to sequence the operations for both processing task and retrieving data so that certain constraints can be satisfied and the maximum completion of all the tasks can be minimized. This problem is formally called as heterogeneous data allocation and task scheduling problem (HDATS).

\section{Literature Review} 
\label{LiteratureReview} 
 of scheduling large-scale scientific workflows onto distributed resources where the workflows are data- intensive, requiring large amounts of data storage,
 
Processors and memories have always been a limited and valuable physical resources for large computations which are summarized in \cite{Ravi1970The}. The problem of scheduling large-scale scientific workflows with distributed resources has been identified by \cite{2007Scheduling}. Their work was extended in \cite{2016Distributed} that proposed genetic algorithms to handle the computing tasks. \cite{chen2012online} introduced an online heterogeneous dual-core scheduling algorithm for dynamic workloads with real-time constraints, and carried out a series of extensive experiments to compare different workloads and scheduling algorithms. This problem also appears in sparse direct solvers, as studied by \cite{2012Robust} who analyzed the effect of processor mapping on memory consumption for multi-frontal methods. Based on the research of sparse direct solvers in \cite{Liu1987An}, \cite{2017Dynamic} proposed a heuristic method with problem related knowledge to reduce the minimum peak memory. \cite{zhao2019optimizing} extended the hypergraph partition-based scheduling method and adopted an improved partition technique to alleviate data traffic in distributed data centers.

\cite{du2013efficient} proposed an efficient loop scheduling algorithm to tackle the problem of expensive write operations on non-volatile main memory for chip multiprocessors, which reduced the number of write operations on non-volatile memories, the processing time, and the energy consumption. 
\cite{s2014Bounded} proposed a bounded memory scheduling algorithm for parallel workloads denoted by dynamic task graphs, where an upper bound is imposed on the peak memory of the computing environment. 
\cite{Sergent2016Controlling} studied the combination between a task-based distributed application and a run-time system to control the memory subscription levels during the processing period. Beyond that, \cite{tsai2013optimized} proposed an improved differential evolution algorithm (IDEA) based on the cost and time models to optimize task scheduling and resource allocation on cloud computing environment. \cite{ergu2013analytic} proposed a model for task-oriented resource allocation in a cloud computing environment. where the resource allocation task is ranked by the pairwise comparison matrix technique and the analytic hierarchy process giving the available resources and user preferences. \cite{praveenchandar2021dynamic} presented an improved task scheduling and power minimization approach for efficient dynamic resource allocation method, which combines a prediction mechanism and dynamic resource table updating algorithm.

There are also research of reducing the task scheduling problem in DSP to the flexible job shop scheduling problem. The FJSP is a well-studied combinatorial optimization problem, which was introduced by \cite{brucker1990job} as an extension of the job shop scheduling problem. For the FJSP with makespan criteria, exact approaches were proposed by \cite{ozguven2010mathematical} and \cite{roshanaei2013mathematical}, who developed mixed-integer linear programming (MILP) models. Another MILP was presented in \cite{birgin2014milp} for the FJSP with an extension that allows precedence relations between operations of a job to be given by an arbitrary directed acyclic graph. \cite{hansmann2014flexible} combined a MILP with a branch and bound algorithm to solve the FJSP with restricted machine accessibility. \cite{zhang2017dynamic} proposed a method based on a two-stage strategy to maximize task scheduling performance and minimize non-reasonable task allocation in clouds, where a job classifier motivated by a principle of Bayes classifier and a dynamic match strategy are utilized. For the task scheduling in virtual controllers and multiple clusters of remote radio heads, \cite{xia2019programmable} translated it into a matroid constrained submodular maximization problem and propose heuristic algorithms to find solutions with half approximation. \cite{fu2020optimal} introduced an unified graph to model the map task scheduling and the reduce task scheduling respectively, and transformed the problem to the well-known graph problem: minimum weighted bipartite matching.

In fog computing based on containers for smart manufacturing, \cite{yin2018tasks} built a new task-scheduling model by considering the role of containers, and designed a task-scheduling algorithm and a reallocation mechanism to reduce task delays in accordance with the characteristics of the containers. \cite{yuan2018spatial} proposed a spatial task scheduling and resource optimization method to minimize the total cost of their provider by cost-effectively scheduling all arriving tasks of heterogeneous applications to meet tasks' delay-bound constraints in distributed green cloud data centers. \cite{hu2020reformed} studied the task scheduling problem to minimize the schedule length of parallel applications while satisfying the energy constraints in heterogeneous distributed systems. For the fairness-aware task scheduling and resource allocation in unmanned aerial vehicle-enabled mobile edge computing networks, \cite{zhao2021fairness} proposed iterative algorithm to deal with them in a sequence and a penalty method-based algorithm to reduce computation complexity. \cite{zhuge2012minimizing} introduced a polynomial-time algorithm based on dynamic programming approach, and a global data allocation algorithm, and a heuristic maximal similarity scheduling to reduce memory traffic and minimize the cost of accessing memory. \cite{zuo2013self} proposed a self-adaptive learning particle swarm optimization based scheduling approach for hybrid infrastructure as a service cloud.

For the data allocation and task scheduling on heterogeneous multiprocessor systems, the main purpose is to find a schedule for the tasks and memories to guarantee that at any time during the execution the memory usage does not exceed its maximum capacity. To solve this problem, we propose a tabu search algorithm which combines several distinguished features such as a greedy and random initial solution construction method and a mixed neighborhood evaluation strategy based on exact evaluation and approximate evaluation. Experimental results on randomly generated instances show that the the proposed TS algorithm can obtain relatively good solutions in a reasonable computational time. We also analyze some key features of TS to identify the performance of the tabu search algorithm.

The rest of the paper is organized as follows: Section \ref{sec_mip} describes the problem and its mathematical formulation. Section \ref{Sec_alg} presents the details of the proposed tabu search method and each of its components. Section \ref{sec_exp} reports the computational results and analyze its key features, and Section \ref{sec_concl} concludes the paper and suggests the future research directions.

\section{Problem definition and formulation}
\label{sec_mip}

\subsection{Problem description}
\begin{figure}[!t]
        \centering
        \includegraphics[width=0.5\textwidth]{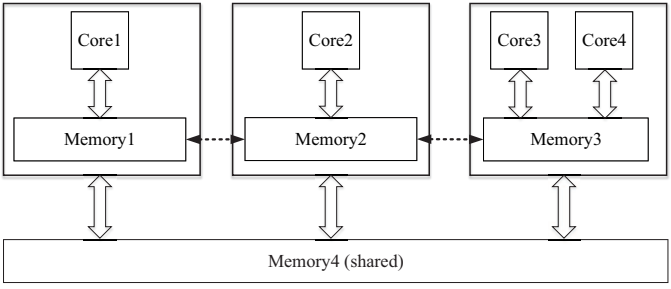}
        \caption{The architecture of the heterogeneous distributed shared-memory multiprocessor system}
        \label{fig_arch}
\end{figure}

The architecture model of the DSP in this paper is a heterogeneous distributed shared-memory multiprocessor system which is described in Fig. \ref{fig_arch}. The architectural scheme encompasses a set $P$ of $n$ connected heterogeneous processors, i.e., $P = \{P_1, P_2, \dots, P_n\}$. Each processor $P_i$ is tightly affiliated with its own local memory $M_i$, and all local memories of the processors constitute a distributed physical memory which are globally shared. For instance, $M_1$ is the local virtual memory of processor $P_1$, while $M_2$ and $M_3$ are remote physical memories. For processor $P_2$, $M_2$ is the local memory, while $M_1$ and $M_3$ are remote memories. Since all  the distributed memories are integrated into a global shared space, every processor has full memory access right to read from or write to a memory. Note that different processors’ accesses operation on the same memory require different times because the structure of memory paradigm is non-uniform.

\begin{figure}[!t]
        \centering
        \includegraphics[width=0.5\textwidth]{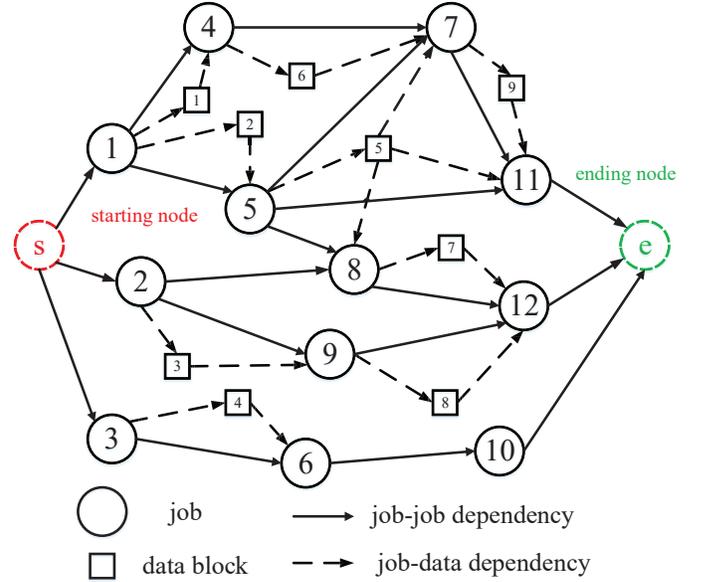}
        \caption{An illustrative example with 12 tasks and 9 data blocks}
        \label{fig_example}
\end{figure}

Given a direct acyclic task graph DAG $G=(V,E)$, $V$ is the node set and $E$ is the edge set, nodes $s,e\in V$ represent the starting and ending nodes, respectively. In the considered problem, by formulating a memory access operation as a node, the traditional DAG can be extended to a memory-access data flow graph (MDFG). Fig. \ref{fig_example} gives an illustrative example of the HDATS problem, where cycle blocks represent the tasks and square blocks represent the data blocks depending on them or being depended.

A MDFG is a node-weighted directed graph extended from a DAG which is described by $G^{'}=(V_1, V_2, \dots, E,D,var,P,M,AT,ET)$, where the notations are explained as follows:
\begin{itemize}
    \item $V_1=\{v_1,V_2,\dots,v_{N_1}\}$ represents a set of $N_1$ task nodes.
    \item $V_2=\{u_1,u_2,\dots,u_{N_2}\}$ represents a set of $N_2$ memory access operation nodes.
    \item $E$ is a set of edges, where $E\subset V\times V$, $V=V_1\cup V_2$. An edge $(i,j) \in E$ denotes the dependency between node $i$ and node $j$, expressing that task or operation $i$ has to be executed before task or operation $j$. 
    \item $D$ is a set of initial input data.
    \item $var: V_1\times V_2\times D \rightarrow \{0,1\}$ is a binary mapping relationship, in which $var(v,u,w)$ represents whether memory access operation $u\in V_2$ is delivering data $w\in D$ for task $v\in V_1$. 
    \item $P=\{P_1,P_2,\dots, P_n\}$ represents a set of $n$ heterogeneous processors.
    \item $M=\{M_1,M_2,\dots,M_m\}$ represents a set of $m$ local memories.
    \item $AT$ is the memory access time functions.
    \item $PT(v_i, P_j) = et_j(i)$ is the processing time of task $v_i$  when it is processed on processor $P_j$.
\end{itemize}

Therefore, the formal definition of the heterogeneous data allocation and task scheduling problem is a MDFG with the aim of seeking a solution denoted by a triple $Mem, AS, SC$, in which $Mem$ is a data allocation $Mem:D\rightarrow M$, where $Mem(h)\in M$ is the memory to store $h\in D$; $AS$ is a task assignment $AS:V_1\rightarrow P$, where $A(v)$ is the processor to execute task $v\in V_1$; $SC$ is a schedule, $SC: V_1\cup V_2\rightarrow \mathbb{R}$, i.e., the starting time of each task in $V_1$ and each memory access operation in $V_2$, such that the amount $T_i$ of data blocks assigned to memory $M_i$ does not exceed its capacity $S(M_i)$, i.e., $T_i<=S(M_i)$, and the total completion time of all the tasks $T(G^{'})$ is minimized. The HDATS problem has been proved to be NP-hard \citep{shao2005efficient}.

\subsection{The integer linear programming formulation of HDATS}

In this section, the integer linear programming (ILP) formulation for HDATS problem is presented, which consists of a task assignment with processor constraint, a data allocation with memory size and concurrency constraints, precedence constraints, a time constraint. Given a MDFG, the ILP model of the HDATS problem encompass two major parts, i.e., a processor assignment and a memory allocation. The processor assignment is to find a task assignment for all tasks of a given MDFG, and the memory allocation is to find a data allocation for all data needed in processing tasks. The objective is to minimize the maximum completion time of all the tasks, i.e.,
\begin{equation}
\label{equ_obj}
\min \max\{RT(i,j)+PT(v_i,P_j)\},  \forall i\in [1,N_{1}],j\in[1,n]
\end{equation}

\subsubsection{Task assignment and processor constraints}

\begin{equation}
\label{equ_t1}
\sum_{j=1}^{n}\sum_{k=1}^{S}x_{ijk}=1, \ \ \ \forall i\in [1,N_{1}]
\end{equation}

\begin{equation}
\label{equ_t2}
\sum_{j=1}^{N_{1}}x^{'}_{ijm}\leq 1, \ \ \ \forall m\in [1,S]
\end{equation}

\begin{equation}
\label{equ_t3}
\sum_{i=1}^{N_{1}}\sum_{j=1}^{n}x^{'}_{ijm}\leq 1, \ \ \ \forall m\in [1,S]
\end{equation}

\begin{equation}
\label{equ_t4}
P(i)=\sum_{j=1}^{n}\sum_{k=1}^{S}j\times x_{ijk} \ \ \ \forall i\in [1,N_{1}]
\end{equation}

\begin{equation}
\label{equ_t5}
x_{ijk}=
	\begin{cases}
	1 \hspace{0.3cm} \text{if task $v_i$ starts to process at stage $k$}\\
	\hspace{0.4cm} \text{ on processor $P_j$,}\\
	0 \hspace{0.3cm} \text{otherwise.}
	\end{cases}
\end{equation}

\begin{equation}
\label{equ_t6}
x^{'}_{ijm}=
	\begin{cases}
	1 \hspace{0.3cm} \text{if task $v_i$ is processed at step $k$}\\
	\hspace{0.5cm} \text{ on processor $P_j$,}\\
	0 \hspace{0.3cm} \text{otherwise.}
	\end{cases}
\end{equation}

In the processor part, let two binary variables $x_{ijk}$ and $x^{'}_{ijm}$ denote whether task $v_i$ in an MDFG $G^{'}$ starts to execute, and is processed in stage $m$ on processor $P_j$, respectively. Constraint (\ref{equ_t1}) ensures that each task node can start execution in one and only one stage and one processor. Constraint (\ref{equ_t2}) ensures that utmost one task is scheduled in any stage on any processor.
Constraint (\ref{equ_t3}) ensures that the number of tasks processed in each stage does not exceed the number of processors.
Formula (\ref{equ_t4}) defines the processor $P(i)$ that is assigned to task $v_i$.

\subsubsection{Data allocation and memory constraints}

\begin{equation}
\label{equ_d1}
\sum_{j=1}^{n}d_{hj}=1 \ \ \ \forall h\in [1,N_{d}]
\end{equation}

\begin{equation}
\label{equ_d2}
\sum_{h=1}^{N_d}d(h)\times d_{hj}\leq S_j \ \ \ \forall j\in [1,n]
\end{equation}

\begin{equation}
\label{equ_d3}
Mem(h)=\sum_{i=1}^{n}j\times d_{hj} \ \ \ \forall h\in [1,N_d]
\end{equation}

\begin{equation}
\label{equ_d4}
\sum_{j=1}^{n}\sum_{k=1}^{S}y_{ljk}= 1, \ \ \ \forall l\in [1,N_2]
\end{equation}

\begin{equation}
\label{equ_d5}
\sum_{l=1}^{N_2}y^{'}_{ljm}\leq MA, \ \ \ \forall j\in [1,n],\forall m\in[1,S]
\end{equation}

\begin{equation}
\label{equ_d6}
M(l)=\sum_{j=1}^{n}\sum_{k=1}^{S}j\times y_{ljk}, \ \ \ \forall l\in [1,N_2]
\end{equation}

\begin{equation}
\label{equ_t7}
	d_{ij}=
	\begin{cases}
	1 \hspace{0.3cm} \text{if data $i$ is allocated to memory $M_j$,}\\
	0 \hspace{0.3cm} \text{otherwise,}
	\end{cases}
\end{equation}

\begin{equation}
\label{equ_t8}
	y_{ljk}=
	\begin{cases}
	1 \hspace{0.3cm} \text{if memory access operation node $u_l$ starts }\\ \hspace{0.3cm}\text{to execute in step $k$ on local memory $M_j$,}\\
	0 \hspace{0.3cm} \text{otherwise.}
	\end{cases}
\end{equation}

\begin{equation}
\label{equ_t9}
	y_{ljk}=
	\begin{cases}
	1 \hspace{0.3cm} \text{if memory access operation node $u_l$ is  }\\ 
	\hspace{0.3cm}\text{scheduled in step $k$ on local memory $M_j$,}\\
	0 \hspace{0.3cm} \text{otherwise.}
	\end{cases}
\end{equation}
In the memory part, let binary variable $d_{ij}$ represents whether data $i$ is allocated to memory $M_j$. Let binary variables $y_{ljk}$ and $y^{'}_{ljk}$ represent whether memory access operation node $u_l$ starts to process, and is scheduled in stage $k$ on memory $M_j$, respectively. Let $Sj$ denotes the capacity of memory $M_j$. Let $M(l)$ be the dependency between data allocation and memory access operations.

Constraint (\ref{equ_d1}) ensures that each data block is allocated to one and only one local memory.
Constraint (\ref{equ_d2}) ensures that the size of all data allocated in $M_j$ is no larger than $S_j$.
Constraint (\ref{equ_d3}) denotes the local memory $Mem(h)$ to store data $h$.
Constraint (\ref{equ_d4}) ensures that each memory access operation node can start processing in one and only one stage and one local memory.
Constraint (\ref{equ_d5}) ensures that the number of memory access operation nodes in each stage does not exceed the access number of a local memory. 
The memory module $M(l)=Mem(D(l))$ for the memory access operation $u_l$ for data $D(l)$ is expressed in constraint (\ref{equ_d6}).

\subsubsection{Precedence constraints}

\begin{equation}
\label{equ_p1}
\begin{split}
\sum_{j=1}^{n}\sum_{k=1}^{S}(k+RT(u,j))\times x_{ujk}\leq \sum_{j=1}^{n}\sum_{k=1}^{S}k\times x_{vjk},\\ \ \ \ \forall e(u,v)\in E, \forall u\in[1,N_1],\forall v\in[1,N_1]
\end{split}
\end{equation}

\begin{equation}
\label{equ_p2}
\begin{split}
\sum_{j=1}^{n}\sum_{k=1}^{S}(k+RT(u,j))\times x_{ujk}\leq \sum_{j=1}^{n}\sum_{k=1}^{S}k\times y_{vjk},\\ \ \ \ \forall e(u,v)\in E, \forall u\in[1,N_1],\forall v\in[1,N_1]
\end{split}
\end{equation}

\begin{equation}
\label{equ_p3}
\begin{split}
\sum_{j=1}^{n}\sum_{k=1}^{S}(k+RA\_t(u,j))\times y_{ujk}\leq \sum_{j=1}^{n}\sum_{k=1}^{S}k\times y_{vjk},\\ \ \ \ \forall e(u,v)\in E, \forall u\in[1,N_1],\forall v\in[1,N_1]
\end{split}
\end{equation}

\begin{equation}
\label{equ_p4}
\begin{split}
\sum_{j=1}^{n}\sum_{k=1}^{S}(k+RA\_t(u,j))\times y_{ujk}\leq \sum_{j=1}^{n}\sum_{k=1}^{S}k\times x_{vjk},\\ \ \ \ \forall e(u,v)\in E, \forall u\in[1,N_1],\forall v\in[1,N_1]
\end{split}
\end{equation}

In a given MDFG, edge $e(u, v)\in E$ denotes the precedence relation from node $u$ to node $v$. Eqs. (\ref{equ_p1})–(\ref{equ_p4}) ensure that each task and memory access operation accurately respect the precedence constraints. Eq. (\ref{equ_p1}) and Eq. (\ref{equ_p3}) respectively formulates the precedence relation among tasks and memory access operations. Eqs. (\ref{equ_p2}) and (\ref{equ_p4}) define the precedence constraints between tasks and memory access operations. Generally, the above equations describe that $u$ must be completed before $v$ can be started.

\subsubsection{Execution and memory access time constraints} \label{subsec_et}

\begin{equation}
\label{equ_e1}
RT(i,j)=\sum_{k=1}^{S}x_{ijk}\times PT(v_i,P_j), \ \ \ \forall i\in[1,N_1],\forall j\in [1,n]
\end{equation}

\begin{equation}
\label{equ_e2}
\sum_{m=1}^{S}x^{'}_{ijm}\leq RT(i,j), \ \ \ \forall i\in[1,N_1],\forall j\in [1,n]
\end{equation}

\begin{equation}
\label{equ_e3}
\sum_{m=k}^{k+RT(i,j)-1}x^{'}_{ijm}= RT(i,j), \ \ \ \forall i\in[1,N_1],\forall j\in [1,n]
\end{equation}

\begin{equation}
\label{equ_e4}
\begin{split}
RT\_t(l,j)=\sum_{i=1}^{N_1}\sum_{h=1}^{N_d}\sum_{k=1}^{S}y_{ljk}var(v_i,u_l,h)AT(P(i),M_)d(h),\\ \ \ \ \forall l\in[1,N_2],\forall j\in [1,n]
\end{split}
\end{equation}

\begin{equation}
\label{equ_e5}
\sum_{m=1}^{S}y^{'}_{ljm}\leq RA\_t(l,j), \ \ \ \forall l\in[1,N_2],\forall j\in [1,n]
\end{equation}

\begin{equation}
\label{equ_e6}
\sum_{m=k}^{k+RT(i,j)-1}y^{'}_{ljm}= RA\_t(l,j), \ \ \ \forall l\in[1,N_2],\forall j\in [1,n].
\end{equation}

In this part, $RT(i, j)$ is the real processing time of task $v_i$ on processor $P_j$ which is defined in Eq. (\ref{equ_e1}).
Constraint (\ref{equ_e2}) presents the relationship that should be satisfied between $x^{'}_{ijm}$ and $RT(i,j)$ for each task. 
If $x_{ijk}=1$, then $x^{'}_{ijm}$ must satisfy the following constraint  (\ref{equ_e3}):
which means the processing of a task should not be interrupted.

Let $RA\_t$ be the real memory access time of a memory access operation $u_l$ on memory $M_j$ which is expressed in constraint (\ref{equ_e4}), 
For each memory access operation, the relationship between $y^{'}_{ljm}$ and $RA\_t(l, j)$ is defined in constraint (\ref{equ_e5}).
which is similar to Eq.(\ref{equ_e2}). If $y_{ijk} = 1$, then $y^{'}_{ijm}$ must satisfy Eq. (\ref{equ_e6}).

\section{Algorithm description}
\label{Sec_alg}
The proposed tabu search algorithm consists of a greedy initial solution construction procedure, the neighbourhood structure, the mixed evaluation strategy, and a tabu search procedure, which are illustrated in details in the following sections.

\subsection{Greedy construction procedure for initial solution}
To efficiently construct a feasible initial solution is primarily important for starting an heuristic algorithm. In this paper, we propose a greedy construction procedure according to the characteristics of the considered problem to generate a feasible solution with high quality in a short time.

The construction of an initial solution is to assign each task to a certain core and each data block to a certain block of memory. However, not all assignments are legal, since the topological relationship between tasks and the capacity constraints on each memory needs to be satisfied. Specifically, some tasks will depend on some other tasks or data blocks. Therefore, the task must wait for all the tasks it depends on to complete, and all the data it depends on to be written  before it can start executing. Besides, when allocating memory for each data block, it is required that the peak memory usage of each block does not exceed its maximum capacity during the entire process. In addition, being limited by types and labels, the candidate cores/memories that are compatible for each task/data block is just a subset of all cores/memories.

The pseudo code of the greedy construction procedure for initial solution is presented in Algorithm \ref{alg_init}, where the main idea can be briefly summarized as iteratively selecting the most important task among the currently unallocated tasks, and then assigning it to the best core and the best memory for the data it produces. 

\begin{algorithm}[!htbp]
\caption{Greedy construction procedure for initial procedure}
\label{alg_init}
{\footnotesize
\begin{algorithmic}[1]
\State\textbf{Input}: Problem instance
\State\textbf{Output}: A feasible initial solution $S_{init}$
\State $S_{init}\gets InitS()$, $taskSet,R,Q,Slack\gets Init()$, $t\gets -1$
\While{$taskSet$ is not empty}
    \State $t\gets selectTaskAccodingToRQSlack()$
    \State $availCores \gets getAvailableCores(t)$;
    \State $endTime \gets InitET(availCores)$
    \For{each core $c$ of in $avalilCores$}
        \State $N\gets getPredecessorsSet(t)$
        \State $startTime \gets \max \{getFinishTime(p)|p\in N\}$
        \For{each data $d$ of task $t$}
            \If{memory of highType2 is enough at startTime}
                \State $tryAssignMemory(d,highType2)$
            \ElsIf{memory highType1 is enough at startTime}
                \State $tryAssignMemory(d,highType1)$
            \Else
                \State $tryAssignMemory(d,lowType)$
            \EndIf
        \EndFor
        \State $endTime[c] \gets calcuEndTime(t,c)$
    \EndFor
    \State $C\gets \arg \min \{getEndTime(c)|c\in avalilCores \} $ \label{alg1_minT}
    \State $assignToCore(t,C),updateSolution(S_{init})$ \label{alg1_ascore}
    \State $freshRQSlack(t,C),freshMemory()$ \label{alg1_rqs}
    \State $taskSet\gets tastSet\setminus\{t\}$ \label{alg1_del}
\EndWhile
\State \Return $S_{init}$
\end{algorithmic}
}
\end{algorithm}

\subsubsection{Preprocessing}
Before starting construction, a preprocessing work is required to generate an profitable job sequence: First, we use the topological sorting to obtain a legal topological sequence that only considers job-job constraints and job-data constraints. Then, we perform dynamic programming procedure on the topological sequence, and calculate the $R$, $Q$, makespan, and $Slack$ values, where $taskSet$ represents the candidate task list that has not been decided yet. Subsequently, in line 5, the task in the front of the candidate list is selected. If there are multiple eligible frontier tasks, we select one of them according to the following lexicographical order:

\begin{enumerate}
\item $R$ value;
\item $Slack$ value;
\item the minimum $Slack$ value of the successor jobs.
\end{enumerate}

Given a direct acyclic task graph DAG $G=(V,E)$, nodes $s,e\in V$ represent the starting and ending nodes, respectively. Let $P_i$ and
$S_i$, be the set with the direct predecessors and successors of
node $i\in V$. Let $R[i]$ and $Q[i]$, be the length of the longest path from starting node $s$ to node $i$, and node $i$ to ending node $e$, respectively, which can be expressed as follows:
\begin{eqnarray}
R[i] &=& \max_{j\in P[i]} \{R[j]+T[j]\}\label{equ_r}\\
Q[i] &=& \max_{j\in S[i]}\{ Q[j] + T[i]\}\label{equ_q}
\end{eqnarray} 
Let $Slack[i]$ be the maximum time allowed to be postponed without deteriorating the maximum completion time of the whole schedule. By using the definitions of $R$ and $Q$, we have:
\begin{equation}
    Slack[i]=C_{max}-R[i]-Q[i]
\end{equation}
where $C_{max}$ denotes the longest length from the starting node $s$ to ending node $e$.

After selecting a task, a set of candidate cores called \emph{availCores} need to be identified to execute the task according to its type, and then we need to allocate specific cores for the task from \emph{availCores}, and allocate memory for the data blocks it generates. In general, a map called \emph{endTime} is established to indicate the end time when the task is assigned to a certain core. It traverses all the cores in the candidate core set of the current task, and then greedily assigns the memory for the task to generate data. Therefore, the entire time period of the current task under the different cores are obtained, and finally select the core and memory assignments that can  complete the task earliest as the assignment result of the task (line \ref{alg1_minT}).

\subsubsection{Greedy construction}

The main steps of the assignment procedure is as follows:
\begin{itemize}

\item For each specific core, we take the maximum warm-up time of all the data it depends on as the warm-up time of the task, and take the sum of the loading time of the data which is calculated by a piecewise function as the loading time of the task.

\item The release time $R$ of the task on the core can be identified as follows:
\begin{itemize}

\item If all predecessor tasks of this task have been assigned with cores, then the end time of its predecessor tasks are known, and the maximum value of the end time of all predecessor tasks is taken as $R$.

\item The current task can start to be executed only when the last task on its core has been finished. By this relationship, we can determine the warm-up time and loading time of the current task.

\item The previous task on the core has been finished and its corresponding data has been moved out, then the task can start to execute.
\end{itemize}

\item It traverses all the tasks according to the release time $R$ of the current task.
\begin{itemize}

\item If a task has been selected and has not been set to be executed, and its end time is earlier than the start time of the current task, it means that the task has been executed. For each data that the task depends on, if all the tasks using the data have been executed, the data can be released, and the release time is the latest execution completion time of all tasks that depend on the data.

\item If a task has been selected and completed, but the data has not been moved out yet, if the completion time of the task is earlier than the start time of the current task, the task can be set to be completed. For all data generated by the task, If the data is not depended on by other tasks, the release time of the data is the move-out time of the task

\end{itemize}

\item The end time of executing the task and moving out data can be calculated. The data blocks generated by the task are sorted according to the minimum slack value of the task that depends on the data block. Memory is allocated for the data blocks in topological order. For each data block generated by the task Data: We first consider global high-speed memory, then consider consider local high-speed memory in the same group as the current core, so as to minimize the warm-up time and move-out time before moving out. For each piece of candidate memory, we calculate how much the memory has been used, and determine whether the current data block can be put in. The warm-up time is the longest warm-up time, and the transport time is the sum of the transport time of each data block of the task.
\end{itemize}

After the assignment of the task is determined, in the dynamic update phase, the task is exactly assigned to the Core and the solution $S_{init}$ is updated (line \ref{alg1_ascore}), and then the memory usage and the $R$, $Q$, and $Slack$ values of the node need to be updated (line \ref{alg1_rqs}).

\begin{itemize}

\item After the assignment scheme is selected in the specific assignment stage, the information of the global solution needs to be updated, and the data block is released and the memory usage is updated according to the start time of the task.

\item In the preprocessing stage, only the execution time of the task is considered. As the tasks are continuously assigned and completed, the corresponding warm-up and move-in and move-out times are also generated. It is necessary to update the $R$, $Q$, and $Slack$ values of the unassigned task for the selection in the next round. \end{itemize}

After the update is completed, the current selected task is deleted from the candidate set (line \ref{alg1_del}), then the next round of assignment is launched to select the next task and allocate memory for it in the same way. When the $taskSet$ is empty, a complete solution is generated. The quality and feasibility of the solution is guaranteed by the greedy construction procedure.

\subsection{The proposed tabu search procedure}
After an initial solution is obtained through the greedy construction algorithm, the solution is further optimized through the tabu search procedure.

Base on the business requirements, we need to allocate each task to a core in its candidate set under the task topology constraints, and at the same time allocate memory for the data blocks under the memory capacity constraints. Considering these two operations simultaneously may lead the large neighborhood size in local search and complicates the evaluation of neighborhood actions as well.

For this reason, this paper presents a two-layer based local search procedure, where the outer layer considers the scheduling of the task sequence on the machine, and the inner layer consider the allocation of memory. If the memory constraints are ignored, the problem can be viewed as the flexible job shop scheduling problem (FJSP). Therefore, the classic neighborhood structures of this problem (N7 and $k$-insertion) proposed by \cite{ding2019two} can be used as the neighborhood action of the outer layer.

The tabu search procedure can be briefly summarized as follows: First, we construct all neighborhood solutions according to the N7 and exchange core neighborhood structures. Then we apply approximate evaluation method on the feasible neighboring solutions, and select the best $K$ neighborhood solutions, and evaluate them accurately. Finally, we select the best solution according to the accurate evaluation results to replace the current one. Note that in order to avoid the revisiting the previous searched areas in a short period, we adopt a tabu table in the local search process, which means the same neighborhood actions will not be executed within a certain tabu period. The pseudo code for the tabu search procedure is given in Algorithm \ref{alg_ts}.
\begin{algorithm}[!htbp]
\caption{The proposed tabu search procedure for HDATS}
\label{alg_ts}
{\footnotesize
\begin{algorithmic}[1]
\State\textbf{Input}: Greedy Solution $S_{init}$, $\lambda$, $\bar T$, $\bar K$
\State\textbf{Output}: The best found solution $S^{*}$
\State $S_{c}\gets S_{init}$, $S^{*}\gets S_{init}$, $N\gets\emptyset$, $Iter\gets 0$, $Duration\gets 0$  
\While{$Iter<\lambda$ and $Duration<T$} 
    \For{each critical task $t$ in $S_{init}$}
        \State $N^{\pi}\gets constructN7(S_{init},t)$ \label{alg_ts_n_pi}
        \State $N^{\alpha}\gets constructChangeCore(S_{init},t)$ \label{alg_ts_n_alpha}
        \State $N\gets N\cup N^{\pi} \cup N^{\alpha}$
        \State $N\gets checkTabuList(N)$\label{alg_ts_n_cktb}
        \If{$N$ is empty}
            \State $S^{'}\gets randomPerturbation(S_{c})$\label{alg_ts_rand}
        \Else
            \State $topkSet\gets selectApproximateTopK(N)$
            \State $S^{'}\gets \arg \min \{getMakespan(S)|S\in topkSet\} $
        \EndIf
        \State add Move($S_{c},S^{'}$) to tabu list \label{alg_ts_add}
        \State $S_{c}\gets S^{'}$\label{alg_ts_repl}
        \State $S^{'}\gets memoryReassign(S^{'})$\label{alg_ts_update_me}
        \If{$getMakespan(S^{'})< getMakespan(S^{*})$}
            \State $S^{*}\gets S^{'}$
            \State $Iter\gets 0$
        \EndIf
    \EndFor
    \State $N\gets\emptyset$; $Iter\gets Iter+1$
    \State $Duration\gets getDuration()$
\EndWhile
\State \Return $S^{*}$
\end{algorithmic}
}
\end{algorithm}

As described in Algorithm \ref{alg_ts}, the input is the initial solution $S_{init}$ constructed by the greedy strategy, the maximum number of unimproved iterations $\lambda$, the maximum number of accurately evaluated solutions $K$ per iteration, and the longest duration $T$ of the search. The output of the tabu search is the best solution $S^*$ found so far.

First, as commonly used the classical FJSP problem \cite{ding2019two}, it is necessary to identify the critical path, critical operations, and critical blocks. Then we adopt the N7 neighborhood structure (called $N^\pi$ here \cite{Gonz2015Scatter} ) and $k$-insertion neighborhood structure (called $N^\alpha$ here \cite{mastrolilli2000effective}), and construct the neighborhood $N^\pi$ and $N^\alpha$ (line \ref{alg_ts_n_pi} and line \ref{alg_ts_n_alpha}), respectively. Let $N$ denote the union of the two neighborhoods, and the solutions in the tabu state are removed (line \ref{alg_ts_n_cktb}). If $N$ is empty, which means all neighborhood actions are in tabu state, then a random perturbation operation is performed on the current solution $S_c$ (line \ref{alg_ts_rand}).

If $N$ is not empty, we approximately evaluate each of the neighborhood solutions, and sort them according to ascending order of the approximate makespan. Then we select the first $K$ solutions and store them in $topkSet$. Since the approximate makespan are often not accurate, it is necessary to accurately evaluate each solution in $topkSet$, calculate its actual makespan, and select the solution $S^{'}$ with the smallest makespan to replace the current solution. After that, this neighborhood move is added to the tabu table (line \ref{alg_ts_add}) and replace the current solution with the neighborhood solution $S^{'}$ (line \ref{alg_ts_repl}).

Since both the $N^\pi$ and $N^\alpha$ neighborhood moves change the job sequences on the machines, it is also necessary to update the memory allocation status of each data block and re-allocate memory for each data block (line \ref{alg_ts_update_me}). For this purpose, we design a memory update algorithm which is described in detail in Section \ref{subsec_mem_update}. 

\subsection{Memory update procedure}
\label{subsec_mem_update}
The time of each task consists of the transfer time and the execution time. The transfer rates of high-speed memory and low-speed memory are not the same. At the same time, high-speed memory has a capacity limit, so the memory allocation strategy of data blocks will affect the final result. In the whole algorithm process, the memory update procedure will be called repeatedly since there are numbers of iterations. Based on the above two reasons, it is required to design an memory update strategy to handle the memory allocation efficiently.

The memory refresh strategy is mainly based on two basic greedy criteria:
\begin{enumerate}
    \item Assign as many blocks of data to fast memory as possible without violating capacity constraints.
    \item Prioritize ``important" data blocks into high-speed memory.
\end{enumerate}

Based on the fact that makespan cannot be optimized without shortening the length of the critical path, we define the data blocks on the critical path as ``critical data blocks" by analogy with the concept of critical blocks in FJSP. The difference is that tasks only appear once in the entire sequence, while the data blocks may be used by multiple tasks, resulting in various transfer times. Therefore, we use the number of moves transferred on the critical path to measure the importance of each data block.

When the memory is updated, the local search has set the task sequence, and when all the memory is placed at a low speed, a complete solution has been generated. Therefore, we can calculate the start time and duration of each stage, the time when each data block enters the memory and the time when it is moved out of the memory. Besides, the critical path and critical task information can be marked.

The required information is calculated and sequenced as follows:
\begin{enumerate}

    \item The topological order of the solution;

    \item The $R$ value is calculated according to the topological order, and the calculation method is as follows:
    \begin{itemize}
        \item The maximum value of the end time of all predecessor tasks (including the predecessor generated by data block dependencies) plus a period of time which depends on the feature of the edge.
        \item The end time of predecessor task on the same control unit. 
        \item The end time of the execution of the predecessor task on the same machine minus the move-in time of the current task. 

The $R$ value of the current task is the maximum value of the above times.
    \end{itemize}
    \item Similarly, the $Q$ value can be calculated as follows:
    \begin{itemize}
        \item The maximum $Q$ value of all predecessor tasks including the successor generated by data block dependencies.
        \item The $Q$ value of the task on the same control unit.
        \item The $Q$ value of the successor task on the same machine minus the move-out time of the current task.

The $Q$ value of the current task is the sum of its processing time and  the maximum value of the above times.

    \end{itemize}
    \item The data block is moved into memory when the task that generates it starts moving in, and is released after all tasks that depend on it have been moved out. Thus we can calculate the lifetime of the data block.
    \item The critical task can be identified as follows: makespan is the maximum sum of $R$ and $Q$ of each task, and all the tasks where the sum of $R$ and $Q$ equals makespan are critical task.
\end{enumerate}
With the global information, the number of occurrences of the data block on the critical path is the number of critical tasks that generate it or depend on it. The data blocks are sorted according to the descending order of their occurrences.

When trying to put important data blocks into high-speed memory, the peak memory usage may exceed the memory capacity. Therefore, a judgement strategy is needed to ensure that memory capacity constraints are met, which is designed as follows: after calculating the lifetime of all data blocks, the memory usage per second can be calculated through the differential array, and then it can be judged whether it exceeds the limit. However, it is time-consuming to obtain the information per second since the size of makespan is often much larger than the number of data block nodes. It is easy to know that the peak of memory usage must occur when the data blocks are put into the memory, so all the time nodes that generate memory usage changes can be discretized first, and then differentiate them into array, thus to determine whether the peak memory usage exceeds the capacity limit.

Algorithm \ref{alg_me} describes the pseudo code of the memory updating procedure. First, it allocates all data blocks to low-speed memory and  initializes the $dataSet$. Then at each iteration, it performs topological sorting on all the tasks and calculate the $R$, $Q$, and $Slack$ values, and sequentially try to allocate the most important data block that has not been assigned to memory.  If the memory usage does not exceed its capacity limit, then it allocates the data block to the memory.

Subsequently, the critical path may be changed because the allocation of a data block is determined. Therefore, in the next round of circulation, it is necessary to recalculate the $R$, $Q$, and $Slack$ values before the next round of iteration, and re-evaluate the importance of the remaining data according to these information. Since the data block is deleted from $dataSet$ every time the memory is allocated for the selected data block, the memory updating procedure is completed when $dataSet$ is empty.
\begin{algorithm}[!htbp]
\caption{The memory updating procedure}
\label{alg_me}
{\footnotesize
\begin{algorithmic}[1]
\State\textbf{Input}: The temp solution $S^{'}$ in Tabu Search, Problem instance $p$
\State\textbf{Output}: The true solution $S^{t}$
\State $S_{'}\gets InitMemory(S_{'})$, $R,Q,Slack\gets Init()$
\State $dataSet\gets getAllData()$, $taskSeq\gets getAllTask()$ 
\While{$dataSet is not empty$}
    \State $topoSeq\gets TopoSort(taskSeq, p)$
    \State $calcuRQSlack(topoSeq,p,S^{'})$
    \State $D\gets -1$, $maxUseT\gets 0$
    \For{each data $d$ in $dataSet$}
        \State $criticalUse\gets countCriIn(d)+countCriOut(d)$
        \If{$criticalUse > maxUseT$}
            \State $maxUseT \gets criticalUse$
            \State $D \gets d$
        \EndIf
    \EndFor
    \If{memory of highType2 is enough}
        \State $AssignToMemory(d,highType2)$
    \ElsIf{memory highType1 is enough}
        \State $AssignToMemory(d,highType1)$
    \Else
        \State $AssignToMemory(d,lowType)$
    \EndIf
    \State $updataSolution(S^{t})$
    \State $dataSet\gets dataSet\setminus\{d\}$
\EndWhile
\State \Return $S^{t}$
\end{algorithmic}
}
\end{algorithm}

\section{Experiment design and analysis}
\label{sec_exp}
\subsection{Parameter settings and experimental protocol}
\label{subsec_proto}
In subsequent sections we conduct extensive experiments to evaluate the performance of the proposed TS algorithm on four sets of randomly generated instances which named randomCaseA, randomCaseB, randomCaseC and randomCaseD.Each of them has 10 instances. The number of jobs in each instance is 500-1000,which is obtained from actual production examples randomly.We coded TS algorithm in C++ and ran it on a cluster of Intel(R) Xeon(R) CPU E7-8870 @ 2.40Ghz. Table \ref{table_para_set} gives the descriptions and settings of the parameters used in TS where the last column denotes the settings for the set of all the instances. These parameter values are determined by extensive preliminary experiments.

	\begin{table}[!t]
		\centering
		\caption{Parameter settings in TS}\label{table_para_set}
		  \begin{threeparttable}
		\begin{tabular}{p{0.8cm}p{3.6cm}p{2.6cm}}
			\toprule
			Para.  &Description & Value \\
			\midrule
			$K_{max}$      &maximum accurate evaluation    &100\\
			$p$         &memory update round&100\\
			$\theta_1$         &tabu tenure for $N^k$&$m+rand()\%(2*m)$\\
			$\theta_2$         &tabu tenure for $N^{\pi}$&$n+rand()\%n$\\
			$\lambda$    &depth of tabu search &100000\\
			$T_{max}$    &maximum run time of TS &600 seconds \\
			\bottomrule
		\end{tabular}
		 \end{threeparttable}
	\end{table}

\begin{table}[htbp]
  \centering
  \caption{The Basic Information for the Benchmarks}
    \begin{tabular}{p{1.0cm}p{3.0cm}p{3.0cm}}
    \toprule
         Item & Description & Value \\
    \midrule
    \multirow{4}[2]{*}{DAG} 
          & Num. of tasks & [200,300] \\
          & Num. of data blocks & [500,700] \\
          & Num. of cores & 2 high-speed + 8 general \\
          & Num. edges : Num. tasks & 8:1 \\
    \midrule
    \multirow{2}[1]{*}{time} & $T_{in}:T_{proc}:T_{out}$ & 7:15:5 \\
          & $S_{high}:S_{low}$ & 1.2:1 \\
    \midrule
       data   & data size & [1,15000] \\
    \bottomrule
    \end{tabular}%
  \label{tab_ins}%
\end{table}%

Table \ref{tab_ins} presents the basic information of the instances, where columns $T_{in}$, $T_{proc}$, and $T_{out}$  denote to move-in time from global memory before execution, the processing time, and the move-out time from global memory after execution of a task, respectively. $S_{high}$ and $S_{low}$ denote the data access time of high-speed and low-speed memories, respectively. There are 40 instances with different processing time and data access time. Note that all of the instances are with the same DAG and memories sizes, and there are infinite size of low memory size.  
\subsection{The comparison of different initial solution strategy}

The initial solution is iteratively constructed, where each step the most important task is assigned to the best core, and the data block it produces is allocated to the best memory. The criterion of identifying the priority of the tasks is vitally important for the quality of initial solution. There are four metrics for evaluating the importance of the tasks.
  
 \begin{itemize}     
     \item $R$-first strategy
The $R$ value of each task represents the earliest start time of the task. The $R$-first strategy means selecting tasks in the order in which the tasks start, and then assigning core to it and allocating memory to the data blocks it generates. If the $R$ values of the two tasks are the same, then compare their $Slack$ values. If the $Slack$ values are the same, then compare the minimum value of the $Slack$ values of all successor tasks of the incumbent task.

\item $Slack$-first strategy
The $Slack$ value indicates the urgency of the task. A small $Slack$ value of a task indicates that the task has a relatively small active space. In specific, $Slack=0$ means that the task is a critical task and should start to processing once it releases, otherwise it would prolong the makespan. The $Slack$-first strategy is similar to the $R$-first strategy, the only difference is that it hierarchically considers $Slack$ first and then $R$ value.

\item Random strategy
After obtaining the most cutting-edge node set, it randomly selects one task from the cutting-edge nodes each time, then assigns it to a core and allocates memory for its data blocks.

\item Relaxed $R$-first strategy
Under the $R$-first strategy, if there exists small difference between the $R$ values and larger difference between the $Slack$ values of the two tasks, the task with the slightly smaller $R$ value and larger $Slack$ value will be selected first, while the other one with smaller $Slack$ value is abandoned. To avoid missing the task with good attribute, we relax the $R$ value and consider two tasks to be approximately the same if the difference between their $R$ values is within a small range, then hierarchically select the task with smaller $Slack$ value.
\end{itemize}

\begingroup
\setlength{\tabcolsep}{4pt} 
\renewcommand{\arraystretch}{0.6} 
\begin{table*}[htbp]\label{table_init}
  \centering
  \caption{Comparison of the makespan obtained by the tabu search algorithm with different initial solution}
    \begin{tabular}{cccccccccccc}
    \toprule
    \multirow{2}[4]{*}{Instance} & \multicolumn{2}{c}{TS (Slack-Frist)} &       & \multicolumn{2}{c}{TS (R-Frist)} &       & \multicolumn{2}{c}{TS (Rand)} &       & \multicolumn{2}{c}{TS (RelaxR)} \\
\cmidrule{2-3}\cmidrule{5-6}\cmidrule{8-9}\cmidrule{11-12}          & $S_0$    & $S^*$    &       & $S_0$    & $S^*$    &       & $S_0$    & $S^*$    &       & $S_0$    & $S^*$ \\
    \midrule
    \textit{randomCaseA1} & 606074 & 348413 &       & 362823 & 339524 &       & 470576 & 338980 &       & 376195 & 337985 \\
          &       &       &       &       &       &       &       &       &       &       &  \\
    \textit{randomCaseA2} & 831278 & 466610 &       & 542076 & 467090 &       & 672760 & 475377 &       & 540276 & 473230 \\
          &       &       &       &       &       &       &       &       &       &       &  \\
    \textit{randomCaseA3} & 1133081 & 606081 &       & 690344 & 613336 &       & 882980 & 610114 &       & 708680 & 617578 \\
          &       &       &       &       &       &       &       &       &       &       &  \\
    \textit{randomCaseA4} & 1360052 & 761321 &       & 871184 & 771733 &       & 1087429 & 753663 &       & 898110 & 760910 \\
          &       &       &       &       &       &       &       &       &       &       &  \\
    \textit{randomCaseA5} & 1642066 & 926061 &       & 1047247 & 926277 &       & 1344884 & 934052 &       & 1067851 & 927152 \\
          &       &       &       &       &       &       &       &       &       &       &  \\
    \textit{randomCaseA6} & 451932 & 284280 &       & 307594 & 285878 &       & 382830 & 284506 &       & 308091 & 285450 \\
          &       &       &       &       &       &       &       &       &       &       &  \\
    \textit{randomCaseA7} & 701007 & 394816 &       & 446356 & 400293 &       & 561346 & 396292 &       & 447646 & 399976 \\
          &       &       &       &       &       &       &       &       &       &       &  \\
    \textit{randomCaseA8} & 990970 & 534694 &       & 637073 & 533002 &       & 764315 & 533981 &       & 624604 & 533351 \\
          &       &       &       &       &       &       &       &       &       &       &  \\
    \textit{randomCaseA9} & 1235485 & 682684 &       & 783087 & 686142 &       & 999947 & 682065 &       & 793142 & 685679 \\
          &       &       &       &       &       &       &       &       &       &       &  \\
    \textit{randomCaseA10} & 1520976 & 842516 &       & 953306 & 847068 &       & 1217143 & 842005 &       & 949004 & 858474 \\
    \midrule
          &       &       &       &       &       &       &       &       &       &       &  \\
    \textit{Avg.} & 1047292 & 584747.6 &       & 664109 & 587034.3 &       & 838421 & 585103.5 &       & 671359.9 & 587978.5 \\
    \bottomrule
    \end{tabular}%
\end{table*}%
\endgroup

Table \ref{table_init} reports the results of the tabu search algorithm with different initial solutions based on the above four different strategies, which are denoted by Slack-First, R-First, Rand, and RelaxR, respectively. Columns $S_0$ and $S^*$ denote makespan of the initial solution and best found solution obtained by the algorithms, respectively. One observes from Table \ref{table_init} that although the initial solution generated with the Slack-First strategy is the worst, the final solution obtained by TS with the Slack-First strategy is best since it obtains the smallest average makespan of 584747.6. Besides, compared with TS with RelaxR, TS with Slack-First improves the makespan of the final solution by 0.55\%. This indicates that the initial solution has impact with effectiveness of the tabu search algorithm.

\subsection{Implementation of load balancing algorithms and comparison}
Load balancing algorithm is a general task scheduling method in cloud computing which basically balances the load to achieve higher throughput and better resource utilization \cite{Gupta2017Load}. We use a load balancing algorithm as a benchmark method to illustrate the effectiveness of the proposed tabu search algorithm. In the load balancing algorithm, it always selects the task that can start earliest, and sort them on the machine according to the ascending order of the earliest time that can start to move. Besides, it always selects the the most idle core.

\begin{table*}[htbp]
\label{table_ls_lb}
  \centering
  \caption{Comparison between TS and LB on ten randomCases under different memory limit and core numbers}
    \begin{tabular}{ccccccccccc}
    \toprule
    \multicolumn{1}{c}{\multirow{2}[4]{*}{Instance}} & \multirow{2}[4]{*}{Alg.} & \multicolumn{4}{c}{HighSpeedMemory-20\%} &       & \multicolumn{4}{c}{HighSpeedMemory-100\%} \\
\cmidrule{3-11}          &       & H:2/L:2 & H:2/L:4 & H:2/L:6 & H:2/L:8 &       & H:2/L:2 & H:2/L:4 & H:2/L:6 & H:2/L:8 \\
    \midrule
    \multirow{3}[1]{*}{\textit{randomCaseB1}} & LB    & 1432933 & 734099 & 518719 & 425040 &       & 1404754 & 711876 & 495737 & 422815 \\
          & TS    & 1313619 & 660178 & 447149 & 336574 &       & 1236491 & 620701 & 412247 & 313958 \\
          & Ratio & 8.33\% & 10.07\% & 13.80\% & 20.81\% &       & 11.98\% & 12.81\% & 16.84\% & 25.75\% \\
          &       &       &       &       &       &       &       &       &       &  \\
    \multirow{3}[0]{*}{\textit{randomCaseB2}} & LB    & 2060452 & 1053167 & 738849 & 588060 &       & 2021406 & 1013852 & 752311 & 552902 \\
          & TS    & 1878409 & 928423 & 618708 & 459974 &       & 1814555 & 893656 & 593398 & 445119 \\
          & Ratio & 8.84\% & 11.84\% & 16.26\% & 21.78\% &       & 10.23\% & 11.86\% & 21.12\% & 19.49\% \\
          &       &       &       &       &       &       &       &       &       &  \\
    \multirow{3}[0]{*}{\textit{randomCaseB3}} & LB    & 2719737 & 1388044 & 994029 & 731136 &       & 2652843 & 1363367 & 983305 & 710956 \\
          & TS    & 2471825 & 1219171 & 809493 & 604653 &       & 2417181 & 1199263 & 795777 & 593234 \\
          & Ratio & 9.12\% & 12.17\% & 18.56\% & 17.30\% &       & 8.88\% & 12.04\% & 19.07\% & 16.56\% \\
          &       &       &       &       &       &       &       &       &       &  \\
    \multirow{3}[0]{*}{\textit{randomCaseB4}} & LB    & 3361614 & 1709825 & 1235748 & 943059 &       & 3298899 & 1777821 & 1152681 & 939897 \\
          & TS    & 3113048 & 1539651 & 1021411 & 768652 &       & 3063450 & 1519906 & 1004531 & 753172 \\
          & Ratio & 7.39\% & 9.95\% & 17.34\% & 18.49\% &       & 7.14\% & 14.51\% & 12.85\% & 19.87\% \\
          &       &       &       &       &       &       &       &       &       &  \\
    \multirow{3}[0]{*}{\textit{randomCaseB5}} & LB    & 4064859 & 2054323 & 1420240 & 1092212 &       & 4012397 & 2110902 & 1408303 & 1133168 \\
          & TS    & 3810226 & 1858816 & 1250750 & 929974 &       & 3750378 & 1854673 & 1231774 & 922303 \\
          & Ratio & 6.26\% & 9.52\% & 11.93\% & 14.85\% &       & 6.53\% & 12.14\% & 12.53\% & 18.61\% \\
          &       &       &       &       &       &       &       &       &       &  \\
    \multirow{3}[0]{*}{\textit{randomCaseB6}} & LB    & 1183168 & 610893 & 415408 & 350052 &       & 1093761 & 584773 & 416284 & 328006 \\
          & TS    & 1092502 & 553014 & 384601 & 286620 &       & 1021168 & 506650 & 340688 & 259167 \\
          & Ratio & 7.66\% & 9.47\% & 7.42\% & 18.12\% &       & 6.64\% & 13.36\% & 18.16\% & 20.99\% \\
          &       &       &       &       &       &       &       &       &       &  \\
    \multirow{3}[0]{*}{\textit{randomCaseB7}} & LB    & 1749517 & 886106 & 612634 & 486318 &       & 1679404 & 878062 & 592738 & 495505 \\
          & TS    & 1572941 & 777143 & 528727 & 393811 &       & 1515265 & 752568 & 492704 & 375646 \\
          & Ratio & 10.09\% & 12.30\% & 13.70\% & 19.02\% &       & 9.77\% & 14.29\% & 16.88\% & 24.19\% \\
          &       &       &       &       &       &       &       &       &       &  \\
    \multirow{3}[0]{*}{\textit{randomCaseB8}} & LB    & 2376474 & 1252518 & 833821 & 658138 &       & 2332810 & 1216888 & 857730 & 656384 \\
          & TS    & 2167320 & 1067950 & 714538 & 548328 &       & 2118525 & 1038375 & 688568 & 520437 \\
          & Ratio & 8.80\% & 14.74\% & 14.31\% & 16.68\% &       & 9.19\% & 14.67\% & 19.72\% & 20.71\% \\
          &       &       &       &       &       &       &       &       &       &  \\
    \multirow{3}[0]{*}{\textit{randomCaseB9}} & LB    & 3006076 & 1593192 & 1100513 & 889751 &       & 2968040 & 1571656 & 1055565 & 858396 \\
          & TS    & 2802003 & 1377207 & 934240 & 687738 &       & 2735798 & 1358655 & 891153 & 668286 \\
          & Ratio & 6.79\% & 13.56\% & 15.11\% & 22.70\% &       & 7.82\% & 13.55\% & 15.58\% & 22.15\% \\
          &       &       &       &       &       &       &       &       &       &  \\
    \multirow{3}[1]{*}{\textit{randomCaseB10}} & LB    & 3674225 & 1896471 & 1392596 & 963473 &       & 3630897 & 1855778 & 1285530 & 1006734 \\
          & TS    & 3455123 & 1705252 & 1134921 & 847893 &       & 3394879 & 1672559 & 1118196 & 837163 \\
          & Ratio & 5.96\% & 10.08\% & 18.50\% & 12.00\% &       & 6.50\% & 9.87\% & 13.02\% & 16.84\% \\
    \bottomrule
    \end{tabular}%
\end{table*}%
Table \ref{table_ls_lb} presents the results of the proposed tabu search algorithm and the load balancing algorithm (denoted by LB) on 10 randomly generated instances. According to the ratio of high speed memory in the whole memory, it consists of two parts: 20\% and 100\%  of high speed memory. In column H:$x$/L:$y$, $x$ and $y$ denote the number of high-speed cores and general low-speed DSP cores, respectively. The results in Table \ref{table_ls_lb} show that TS improves the makespan by 5.96-25.75\% compared with LB for all the tested instances, which demonstrates the effectiveness of the proposed tabu search algorithm.

\subsection{The stability of the tabu search}
\begin{figure}[!t]
        \centering
        \includegraphics[width=0.5\textwidth]{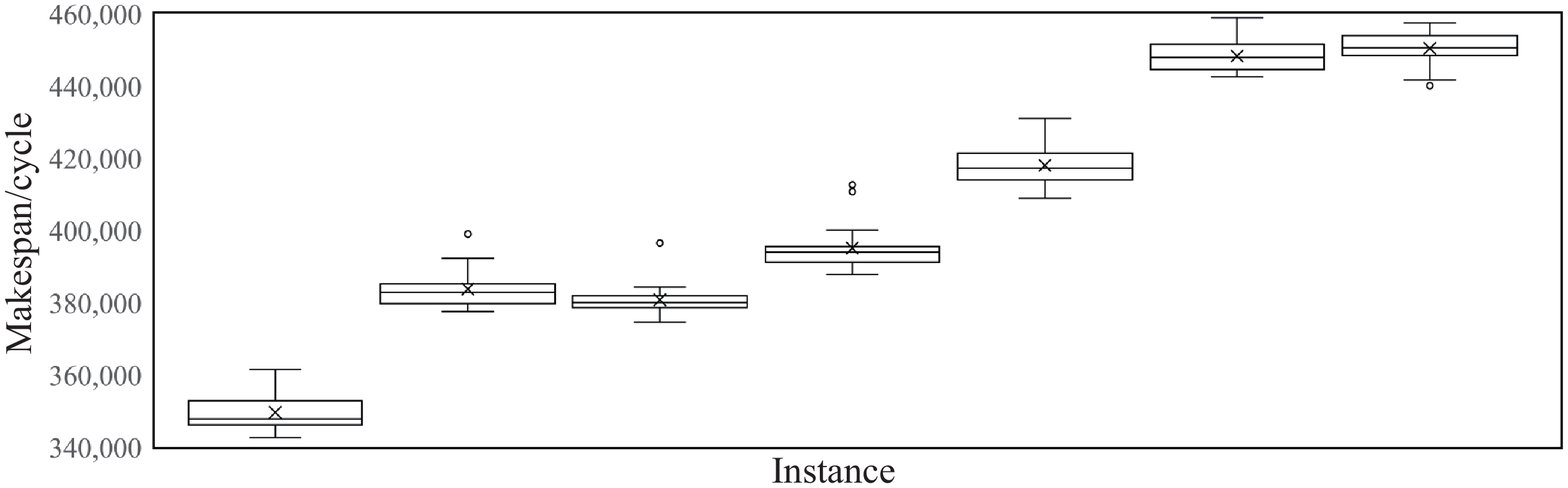}
        \caption{The boxplot of makespan obtained by TS on 10 instances  }
        \label{fig_stability}
\end{figure}

In this section, we analyze the stability of tabu search algorithm by run TS on 10 instances from \emph{randomCaseC1} to \emph{randomCaseC10}. For this purpose, we apply TS on each instances for 20 independent runs, and each run is equipped with a different initial solution. The aim is to detect the difference on the quality of best found solutions. The computational results are plotted in Fig. \ref{fig_stability}. It can be observed that the range of makespan and its mean values are relatively low for each instances, and the difference between the minimum and maximum makespan is almost the same among these instances, which confirms the stability of the proposed tabu search algorithm.

\subsection{The impact of the number of cores}
\begin{figure}[!t]
        \centering
        \includegraphics[width=0.5\textwidth]{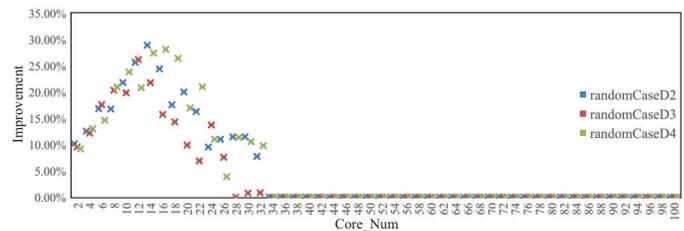}
        \caption{The change of makespan with respect to the number of high speed cores}
        \label{fig_chanage_mk_memory}
\end{figure}

\begingroup
\setlength{\tabcolsep}{2pt} 
\renewcommand{\arraystretch}{1.5} 
\begin{sidewaystable*}[htbp]\tiny
  \centering
  \caption{Comparison between TS and LB with respect to different number of high speed cores}
    \begin{tabular}{ccccccccccccccccccccccccccccccc}
    \toprule
    \multirow{2}[4]{*}{Num.} & \multicolumn{3}{c}{randomCaseD1} & \multicolumn{3}{c}{randomCaseD2} & \multicolumn{3}{c}{randomCaseD3} & \multicolumn{3}{c}{randomCaseD4} & \multicolumn{3}{c}{randomCaseD5} & \multicolumn{3}{c}{randomCaseD6} & \multicolumn{3}{c}{randomCaseD7} & \multicolumn{3}{c}{randomCaseD8} & \multicolumn{3}{c}{randomCaseD9} & \multicolumn{3}{c}{randomCaseD10} \\
\cmidrule{2-31}          & LB    & TS    & Imp. & LB    & TS    & Imp. & LB    & TS    & Imp. & LB    & TS    & Imp. & LB    & TS    & Imp. & LB    & TS    & Imp. & LB    & TS    & Imp. & LB    & TS    & Imp. & LB    & TS    & Imp. & LB    & TS    & Imp. \\
        \midrule
    2     & 3198633 & 2888124 & 9.71\% & 2946630 & 2646980 & 10.17\% & 3056492 & 2764396 & 9.56\% & 3299728 & 2996062 & 9.20\% & 3386055 & 3091910 & 8.69\% & 3549998 & 3276378 & 7.71\% & 3322715 & 3052583 & 8.13\% & 3530529 & 3243464 & 8.13\% & 3745311 & 3414566 & 8.83\% & 3798137 & 3476018 & 8.48\% \\
    4     & 1613162 & 1435472 & 11.02\% & 1494957 & 1307182 & 12.56\% & 1565101 & 1374361 & 12.19\% & 1705887 & 1483618 & 13.03\% & 1738912 & 1516734 & 12.78\% & 1814678 & 1616269 & 10.93\% & 1693341 & 1488885 & 12.07\% & 1871944 & 1587513 & 15.19\% & 1876515 & 1673427 & 10.82\% & 1969425 & 1695215 & 13.92\% \\
    6     & 1121793 & 950900 & 15.23\% & 1038512 & 863557 & 16.85\% & 1098781 & 905511 & 17.59\% & 1164225 & 993670 & 14.65\% & 1224088 & 1010947 & 17.41\% & 1239864 & 1081534 & 12.77\% & 1186216 & 995118 & 16.11\% & 1282006 & 1062742 & 17.10\% & 1307113 & 1116657 & 14.57\% & 1333648 & 1133991 & 14.97\% \\
    8     & 879162 & 707819 & 19.49\% & 779266 & 648545 & 16.77\% & 866031 & 689727 & 20.36\% & 956007 & 755795 & 20.94\% & 945951 & 767166 & 18.90\% & 999844 & 809962 & 18.99\% & 935274 & 748828 & 19.93\% & 1064968 & 805252 & 24.39\% & 995450 & 837599 & 15.86\% & 1065546 & 856122 & 19.65\% \\
    10    & 751317 & 569169 & 24.24\% & 669215 & 523192 & 21.82\% & 695549 & 557363 & 19.87\% & 808228 & 615656 & 23.83\% & 820338 & 619690 & 24.46\% & 806663 & 652603 & 19.10\% & 777568 & 605441 & 22.14\% & 917611 & 652602 & 28.88\% & 832829 & 673674 & 19.11\% & 911366 & 694627 & 23.78\% \\
    12    & 638930 & 482436 & 24.49\% & 597601 & 444264 & 25.66\% & 640345 & 472777 & 26.17\% & 655687 & 519087 & 20.83\% & 663402 & 527766 & 20.45\% & 675623 & 550161 & 18.57\% & 772425 & 554246 & 28.25\% & 776926 & 602834 & 22.41\% & 700731 & 567047 & 19.08\% & 789158 & 601287 & 23.81\% \\
    14    & 604559 & 458431 & 24.17\% & 553472 & 393466 & 28.91\% & 593507 & 464112 & 21.80\% & 631971 & 458717 & 27.41\% & 666627 & 517773 & 22.33\% & 624856 & 476378 & 23.76\% & 743434 & 554246 & 25.45\% & 748838 & 602834 & 19.50\% & 622223 & 490909 & 21.10\% & 770361 & 601287 & 21.95\% \\
    16    & 551788 & 458431 & 16.92\% & 511163 & 386340 & 24.42\% & 550847 & 464112 & 15.75\% & 596871 & 429031 & 28.12\% & 670277 & 517773 & 22.75\% & 582665 & 426931 & 26.73\% & 585693 & 554246 & 5.37\% & 684248 & 602834 & 11.90\% & 536940 & 434687 & 19.04\% & 729197 & 601287 & 17.54\% \\
    18    & 551336 & 458431 & 16.85\% & 468664 & 386340 & 17.57\% & 541649 & 464112 & 14.31\% & 583086 & 429031 & 26.42\% & 613091 & 517773 & 15.55\% & 525365 & 399794 & 23.90\% & 597721 & 554246 & 7.27\% & 668350 & 602834 & 9.80\% & 538069 & 414056 & 23.05\% & 760911 & 601287 & 20.98\% \\
    20    & 604093 & 458431 & 24.11\% & 483166 & 386340 & 20.04\% & 515153 & 464112 & 9.91\% & 516870 & 429031 & 16.99\% & 618600 & 517773 & 16.30\% & 492935 & 399794 & 18.90\% & 628480 & 554246 & 11.81\% & 639094 & 602834 & 5.67\% & 536710 & 414056 & 22.85\% & 709888 & 601287 & 15.30\% \\
    22    & 519229 & 458431 & 11.71\% & 461574 & 386340 & 16.30\% & 498533 & 464112 & 6.90\% & 543043 & 429031 & 21.00\% & 588662 & 517773 & 12.04\% & 484540 & 399794 & 17.49\% & 634449 & 554246 & 12.64\% & 620872 & 602834 & 2.91\% & 486252 & 414056 & 14.85\% & 736865 & 601287 & 18.40\% \\
    24    & 503462 & 458431 & 8.94\% & 427049 & 386340 & 9.53\% & 538027 & 464112 & 13.74\% & 482281 & 429031 & 11.04\% & 579522 & 517773 & 10.66\% & 456069 & 399794 & 12.34\% & 605923 & 554246 & 8.53\% & 668267 & 602834 & 9.79\% & 481680 & 414056 & 14.04\% & 679407 & 601287 & 11.50\% \\
    26    & 487449 & 458431 & 5.95\% & 434194 & 386340 & 11.02\% & 502194 & 464112 & 7.58\% & 446363 & 429031 & 3.88\% & 580815 & 517773 & 10.85\% & 469477 & 399794 & 14.84\% & 595677 & 554246 & 6.96\% & 644709 & 602834 & 6.50\% & 486778 & 414056 & 14.94\% & 630577 & 601287 & 4.64\% \\
    28    & 481586 & 458431 & 4.81\% & 436479 & 386340 & 11.49\% & 464112 & 464112 & 0.00\% & 484036 & 429031 & 11.36\% & 580815 & 517773 & 10.85\% & 429604 & 399794 & 6.94\% & 596567 & 554246 & 7.09\% & 602834 & 602834 & 0.00\% & 474013 & 414056 & 12.65\% & 701373 & 601287 & 14.27\% \\
    30    & 470795 & 458431 & 2.63\% & 436479 & 386340 & 11.49\% & 467758 & 464112 & 0.78\% & 479707 & 429031 & 10.56\% & 517773 & 517773 & 0.00\% & 461356 & 399794 & 13.34\% & 596361 & 554246 & 7.06\% & 636737 & 602834 & 5.32\% & 470519 & 414056 & 12.00\% & 652976 & 601287 & 7.92\% \\
    32    & 458431 & 458431 & 0.00\% & 418867 & 386340 & 7.77\% & 467984 & 464112 & 0.83\% & 475577 & 429031 & 9.79\% & 548791 & 517773 & 5.65\% & 447653 & 399794 & 10.69\% & 554246 & 554246 & 0.00\% & 622928 & 602834 & 3.23\% & 446039 & 414056 & 7.17\% & 659902 & 601287 & 8.88\% \\
    34    & 492587 & 458431 & 6.93\% & 386340 & 386340 & 0.00\% & 464112 & 464112 & 0.00\% & 429031 & 429031 & 0.00\% & 544831 & 517773 & 4.97\% & 438891 & 399794 & 8.91\% & 554246 & 554246 & 0.00\% & 602834 & 602834 & 0.00\% & 445778 & 414056 & 7.12\% & 659902 & 601287 & 8.88\% \\
    36    & 501445 & 458431 & 8.58\% & 386340 & 386340 & 0.00\% & 464112 & 464112 & 0.00\% & 429031 & 429031 & 0.00\% & 547426 & 517773 & 5.42\% & 406142 & 399794 & 1.56\% & 554246 & 554246 & 0.00\% & 602834 & 602834 & 0.00\% & 450462 & 414056 & 8.08\% & 659902 & 601287 & 8.88\% \\
    38    & 473452 & 458431 & 3.17\% & 386340 & 386340 & 0.00\% & 464112 & 464112 & 0.00\% & 429031 & 429031 & 0.00\% & 519071 & 517773 & 0.25\% & 399794 & 399794 & 0.00\% & 554246 & 554246 & 0.00\% & 602834 & 602834 & 0.00\% & 414056 & 414056 & 0.00\% & 605157 & 601287 & 0.64\% \\
    40    & 458431 & 458431 & 0.00\% & 386340 & 386340 & 0.00\% & 464112 & 464112 & 0.00\% & 429031 & 429031 & 0.00\% & 532116 & 517773 & 2.70\% & 449778 & 399794 & 11.11\% & 554246 & 554246 & 0.00\% & 602834 & 602834 & 0.00\% & 414056 & 414056 & 0.00\% & 620827 & 601287 & 3.15\% \\
    42    & 472655 & 458431 & 3.01\% & 386340 & 386340 & 0.00\% & 464112 & 464112 & 0.00\% & 429031 & 429031 & 0.00\% & 537749 & 517773 & 3.71\% & 441016 & 399794 & 9.35\% & 554246 & 554246 & 0.00\% & 602834 & 602834 & 0.00\% & 414056 & 414056 & 0.00\% & 601287 & 601287 & 0.00\% \\
    44    & 458431 & 458431 & 0.00\% & 386340 & 386340 & 0.00\% & 464112 & 464112 & 0.00\% & 429031 & 429031 & 0.00\% & 517773 & 517773 & 0.00\% & 399794 & 399794 & 0.00\% & 554246 & 554246 & 0.00\% & 602834 & 602834 & 0.00\% & 414056 & 414056 & 0.00\% & 601287 & 601287 & 0.00\% \\
    46    & 466430 & 458431 & 1.71\% & 386340 & 386340 & 0.00\% & 464112 & 464112 & 0.00\% & 429031 & 429031 & 0.00\% & 517773 & 517773 & 0.00\% & 399794 & 399794 & 0.00\% & 554246 & 554246 & 0.00\% & 602834 & 602834 & 0.00\% & 414056 & 414056 & 0.00\% & 601287 & 601287 & 0.00\% \\
    48    & 458431 & 458431 & 0.00\% & 386340 & 386340 & 0.00\% & 464112 & 464112 & 0.00\% & 429031 & 429031 & 0.00\% & 517773 & 517773 & 0.00\% & 399794 & 399794 & 0.00\% & 554246 & 554246 & 0.00\% & 602834 & 602834 & 0.00\% & 414056 & 414056 & 0.00\% & 601287 & 601287 & 0.00\% \\
    50    & 458431 & 458431 & 0.00\% & 386340 & 386340 & 0.00\% & 464112 & 464112 & 0.00\% & 429031 & 429031 & 0.00\% & 517773 & 517773 & 0.00\% & 399794 & 399794 & 0.00\% & 554246 & 554246 & 0.00\% & 602834 & 602834 & 0.00\% & 414056 & 414056 & 0.00\% & 601287 & 601287 & 0.00\% \\
    \bottomrule
    \end{tabular}%
  \label{tab_diff_core_num}%
\end{sidewaystable*}%
\endgroup

In this section we analyze the impact of the number of cores to the performance of TS. For this purpose, we apply TS and LB on three instances namely \emph{randomCaseD1} to \emph{randomCaseD3}, and plot the results in Fig. \ref{fig_chanage_mk_memory}, where the $x$-axis represents the number of cores and $y$-axis represents the improvement rate of makespan obtained by TS compared with LB. Note that in order to guarantee the heterogeneity of the architecture, there are at least two synergistic high-speed cores.

From Fig. \ref{fig_chanage_mk_memory}, one observes that the improvement rate increases from 10\% to 30\% when the number of DSP cores increases from 2 to 12, and decreases to 0 when the number of DSP cores increases from 12 to around 28, and always keep at 0 when there are more than 28 DSP cores. The reason lies behind may be that with small number of DSP cores, multiple unrelated tasks are assigned to the same cores and thus leads to that part of them depends on the others, while if there are sufficient cores, the predecessors of one task can be assigned to different cores and can be processed in a parallel way. Table \ref{tab_diff_core_num} reports the detailed results of LB and TS and their differences with 2 to 50 DSP cores.

\subsection{The Effect of mixed evaluation strategy}

It is known to all the the key operations in local search procedure is the evaluation of neighboring solutions. To reduce the computational burden in tabu search, we introduce a mixed evaluation strategy in this paper. In specific, at each iteration of TS, we first apply approximate evaluation method on all the neighboring solutions, which may calculate makespan not that accurately but it runs very fast. Then, we sort these neighboring solutions according to the ascending order of makespan. Subsequently, we apply exact evaluation method the on the top $k$ solutions, and choose the one with minimum makespan to replace the current solution before entering into the next round of tabu search.

\begin{figure}[!t]
        \centering
        \includegraphics[width=0.5\textwidth]{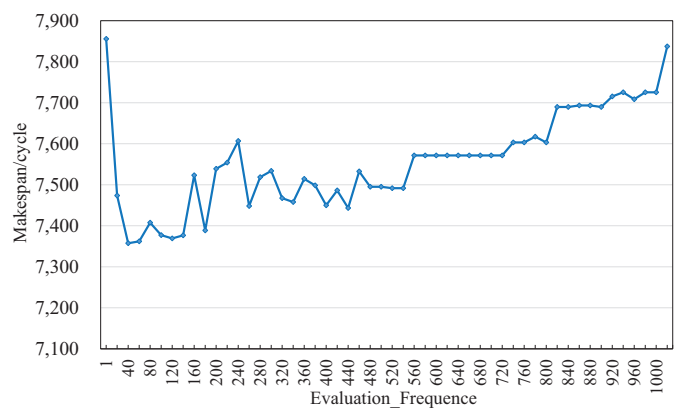}
        \caption{The change of makespan with respect to the ratio of exact evaluation on a random instance}
        \label{fig_mixed_origin}
\end{figure}

\begin{figure}[!t]
        \centering
        \includegraphics[width=0.5\textwidth]{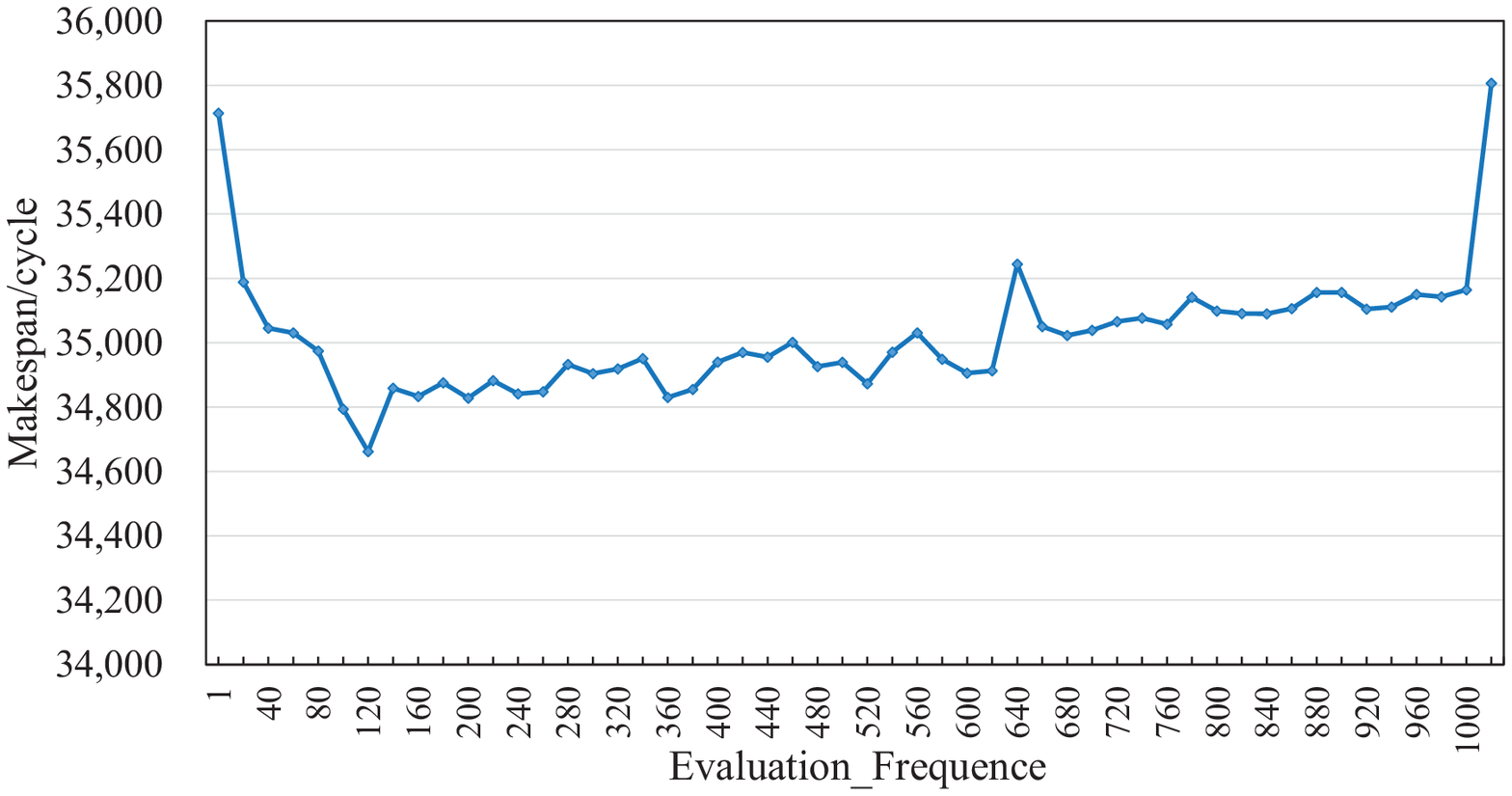}
        \caption{The change of makespan with respect to the ratio of exact evaluation on five times of a random instance}
        \label{fig_mixed_ext}
\end{figure}

Fig. \ref{fig_mixed_origin} and Fig. \ref{fig_mixed_ext} plot the results of TS with respect to the ratio of exact evaluation on a random instance and a larger instance, i.e., 5 times of itself, respectively. One observes that when $k=1$, the left most point in $x$-axis, represents exactly evaluate the best solution measured by approximate method. Both curves decrease when $k\in [1,30]$ and $k\in [1,120]$, and slightly increase with the increase of $k$. This mainly because that, with the same cutoff time, too many runs of exact evaluation strategy leads to the less iterations of tabu search, thus deteriorate its effectiveness. This indicates that the mixed evaluation strategy is a comprise of exact and approximate methods, which can select a relatively high-quality solution quickly. 

\subsection{The effect of high speed memory ratio on makespan}

\begin{figure}[!t]
        \centering
        \includegraphics[width=0.5\textwidth]{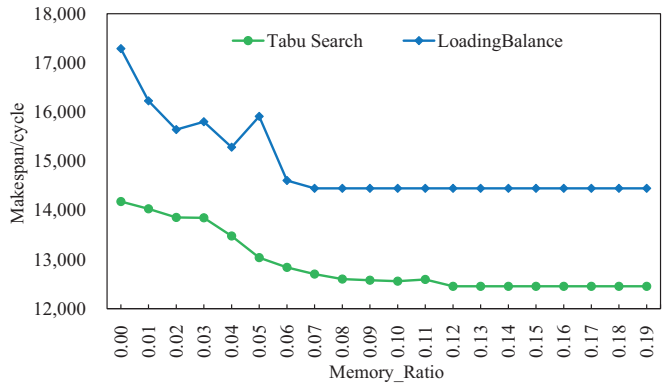}
        \caption{The comparison between TS and LB on different ratio of high speed memory}
        \label{fig_high_memory}
\end{figure}

In this section, we intend to detect the effect of high speed memory ratio on makespan. We apply TS and LB on instance \emph{randomCaseD2} for 20 independent runs respectively, and plot the results in Fig. \ref{fig_high_memory}. One observes that TS outperforms LB for all the memory ratio range from 0 to 0.19 at least with a difference of 2000 in makespan. The reason may be that, due to the greedy strategy adopted by the memory allocation strategy, the two algorithms have an abnormal situation that the proportion of high-speed memory increases slightly, and the makespan increases instead, which has minimum or no impact on the tabu search algorithm.

Besides, when the high-speed memory is insufficient, the makespan obtained by tabu search increases slightly, indicating that the tabu search can more effectively avoid the impact of insufficient high-speed memory. The makespan of tabu search in low speed is still lower than that of load balancing in high speed. Therefore, by limiting the usage of high-speed memory, a better scheduling scheme for both makespan and high-speed memory can be obtained.

\section{Conclusions}
\label{sec_concl}

This paper propose a tabu search algorithm to tackle the task scheduling problem in the digital signal processor. By qualitatively and quantitatively analyzing the performance of load balancing, multi-priority initial solutions, and local search in different cases with different numbers of cores and different high-speed memory ratios, the following conclusions are drawn:

First, the disadvantage of greedy construction mainly occurs in that tasks that are not in a hurry to be executed are assigned resources in the early stage of scheduling, resulting in the remaining tasks with task constraints between each other cannot be parallelized at the end of scheduling, and a large number of resources are idle, resulting in low resource utilization, and the local search can be performed by continuously adjusting the tasks stuck on the critical path due to machine constraints, so that the end time of each machine tends to be consistent.

Second, the local search algorithm has good stability and is little affected by the initial solution goodness and different random seeds.

Third, the tabu search method averagely improves the makespan by 5-25\% compared to load balancing algorithm. Different cores/high-speed memory ratios have an impact on the boost rate, and the load balancing algorithm is unstable and may deteriorate significantly in some cases.

Fourth, hybrid evaluation balances evaluation accuracy and evaluation time, and finally enables the local search to converge to a better solution.

Fifth, the influence of the number of cores on the promotion rate can be regarded as a normalized function, and the number of cores with the maximum promotion rate under different other conditions is not necessarily the same.

Sixth, compared with the greedy algorithm, the local search is more adaptable to the situation of insufficient high-speed memory, and makespan increases less than when the high-speed memory is sufficient, and the scheduling results of the local search are often better than the load on the premise of not using any high-speed memory. 

Future research directions can be combining population-based metaheuristic methods and problem-specific knowledge to enhance the performance of the current algorithm. Besides, Solution-based tabu strategy is also worthy to attempt in order to improve the search intensification of heuristics. Furthermore, another extension to this study could include energy-aware information allocation and task scheduling with the goal of optimizing the total workload to execute and to minimize the total energy consumption.

\bibliographystyle{IEEEtranN}
\bibliography{library,literatureDisA1}

\begin{thebibliography}{42}
\providecommand{\natexlab}[1]{#1}
\providecommand{\url}[1]{#1}
\csname url@samestyle\endcsname
\providecommand{\newblock}{\relax}
\providecommand{\bibinfo}[2]{#2}
\providecommand{\BIBentrySTDinterwordspacing}{\spaceskip=0pt\relax}
\providecommand{\BIBentryALTinterwordstretchfactor}{4}
\providecommand{\BIBentryALTinterwordspacing}{\spaceskip=\fontdimen2\font plus
\BIBentryALTinterwordstretchfactor\fontdimen3\font minus
  \fontdimen4\font\relax}
\providecommand{\BIBforeignlanguage}[2]{{%
\expandafter\ifx\csname l@#1\endcsname\relax
\typeout{** WARNING: IEEEtranN.bst: No hyphenation pattern has been}%
\typeout{** loaded for the language `#1'. Using the pattern for}%
\typeout{** the default language instead.}%
\else
\language=\csname l@#1\endcsname
\fi
#2}}
\providecommand{\BIBdecl}{\relax}
\BIBdecl

\bibitem[Chantem et~al.(2010)Chantem, Hu, and Dick]{chantem2010temperature}
T.~Chantem, X.~S. Hu, and R.~P. Dick, ``Temperature-aware scheduling and
  assignment for hard real-time applications on mpsocs,'' \emph{IEEE
  Transactions on Very Large Scale Integration Systems}, vol.~19, no.~10, pp.
  1884--1897, 2010.

\bibitem[Baruah and Fisher(2006)]{baruah2006partitioned}
S.~Baruah and N.~Fisher, ``The partitioned multiprocessor scheduling of
  deadline-constrained sporadic task systems,'' \emph{IEEE Transactions on
  Computers}, vol.~55, no.~7, pp. 918--923, 2006.

\bibitem[Chen et~al.(2012)Chen, Liao, and Tsai]{chen2012online}
Y.-S. Chen, H.~C. Liao, and T.-H. Tsai, ``Online real-time task scheduling in
  heterogeneous multicore system-on-a-chip,'' \emph{IEEE Transactions on
  Parallel and Distributed Systems}, vol.~24, no.~1, pp. 118--130, 2012.

\bibitem[Chiang et~al.(2006)Chiang, Chang, and Huang]{chiang2006multi}
T.-C. Chiang, P.-Y. Chang, and Y.-M. Huang, ``Multi-processor tasks with
  resource and timing constraints using particle swarm optimization,''
  \emph{IJCSNS International Journal of Computer Science and Network Security},
  vol.~6, no.~4, pp. 71--77, 2006.

\bibitem[Du et~al.(2013)Du, Wang, Zhuge, Hu, and Sha]{du2013efficient}
J.~Du, Y.~Wang, Q.~Zhuge, J.~Hu, and E.~H.-M. Sha, ``Efficient loop scheduling
  for chip multiprocessors with non-volatile main memory,'' \emph{Journal of
  Signal Processing Systems}, vol.~71, no.~3, pp. 261--273, 2013.

\bibitem[Yin et~al.(2018)Yin, Luo, and Luo]{yin2018tasks}
L.~Yin, J.~Luo, and H.~Luo, ``Tasks scheduling and resource allocation in fog
  computing based on containers for smart manufacturing,'' \emph{IEEE
  Transactions on Industrial Informatics}, vol.~14, no.~10, pp. 4712--4721,
  2018.

\bibitem[Ilavarasan and Thambidurai(2007)]{ilavarasan2007low}
E.~Ilavarasan and P.~Thambidurai, ``Low complexity performance effective task
  scheduling algorithm for heterogeneous computing environments,''
  \emph{Journal of Computer sciences}, vol.~3, no.~2, pp. 94--103, 2007.

\bibitem[Kang et~al.(2011)Kang, He, and Song]{kang2011task}
Q.~Kang, H.~He, and H.~Song, ``Task assignment in heterogeneous computing
  systems using an effective iterated greedy algorithm,'' \emph{Journal of
  Systems and Software}, vol.~84, no.~6, pp. 985--992, 2011.

\bibitem[Kang and Dean(2010)]{kang2010darts}
S.~Kang and A.~G. Dean, ``Darts: Techniques and tools for predictably fast
  memory using integrated data allocation and real-time task scheduling,'' in
  \emph{2010 16th IEEE Real-Time and Embedded Technology and Applications
  Symposium}.\hskip 1em plus 0.5em minus 0.4em\relax IEEE, 2010, pp. 333--342.

\bibitem[Lakshmanan et~al.(2009)Lakshmanan, de~Niz, and
  Rajkumar]{lakshmanan2009coordinated}
K.~Lakshmanan, D.~de~Niz, and R.~Rajkumar, ``Coordinated task scheduling,
  allocation and synchronization on multiprocessors,'' in \emph{2009 30th IEEE
  Real-Time Systems Symposium}.\hskip 1em plus 0.5em minus 0.4em\relax IEEE,
  2009, pp. 469--478.

\bibitem[Ouni et~al.(2011)Ouni, Ayadi, and Mtibaa]{ouni2011partitioning}
B.~Ouni, R.~Ayadi, and A.~Mtibaa, ``Partitioning and scheduling technique for
  run time reconfigured systems,'' \emph{International Journal of Computer
  Aided Engineering and Technology}, vol.~3, no.~1, pp. 77--91, 2011.

\bibitem[Wang et~al.(2014)Wang, Li, Chen, He, and Li]{wang2014energy}
Y.~Wang, K.~Li, H.~Chen, L.~He, and K.~Li, ``Energy-aware data allocation and
  task scheduling on heterogeneous multiprocessor systems with time
  constraints,'' \emph{IEEE Transactions on Emerging Topics in Computing},
  vol.~2, no.~2, pp. 134--148, 2014.

\bibitem[Ravi et~al.(1970)Ravi, Sethi, J., D., and Ullman]{Ravi1970The}
Ravi, Sethi, J., D., and Ullman, ``The generation of optimal code for
  arithmetic expressions,'' \emph{Journal of the {ACM}}, 1970.

\bibitem[Ramakrishnan et~al.(2007)Ramakrishnan, Singh, Zhao, Deelman, and
  Samidi]{2007Scheduling}
A.~Ramakrishnan, G.~Singh, H.~Zhao, E.~Deelman, and M.~Samidi, ``Scheduling
  data-intensiveworkflows onto storage-constrained distributed resources,'' in
  \emph{Proceedings of the Seventh IEEE International Symposium on Cluster
  Computing and the Grid}, 2007.

\bibitem[Peris et~al.(2016)Peris, Hernández, and Huedo]{2016Distributed}
A.~D. Peris, J.~Hernández, and E.~Huedo, ``Distributed late-binding scheduling
  and cooperative data caching,'' \emph{Journal of Grid Computing}, 2016.

\bibitem[Rouet et~al.(2012)Rouet, Agullo, Amestoy, Buttari, and
  Excellent]{2012Robust}
F.~H. Rouet, E.~Agullo, P.~R. Amestoy, A.~Buttari, and J.~Y.~L. Excellent,
  ``Robust memory-aware mappings for parallel multifrontal factorizations,''
  \emph{SIAM Journal on Scientific Computing}, vol.~38, no.~3, 2012.

\bibitem[Liu(1987)]{Liu1987An}
J.~Liu, ``An application of generalized tree pebbling to sparse matrix
  factorization,'' \emph{SIAM Journal on Algebraic Discrete Methods}, 1987.

\bibitem[Aupy et~al.(2017)Aupy, Brasseur, and Marchal]{2017Dynamic}
G.~Aupy, C.~Brasseur, and L.~Marchal, ``Dynamic memory-aware task-tree
  scheduling,'' in \emph{Parallel \& Distributed Processing Symposium}, 2017.

\bibitem[Zhao et~al.(2019)Zhao, Yang, Munir, Liu, Li, and
  Qu]{zhao2019optimizing}
L.~Zhao, Y.~Yang, A.~Munir, A.~X. Liu, Y.~Li, and W.~Qu, ``Optimizing
  geo-distributed data analytics with coordinated task scheduling and
  routing,'' \emph{IEEE Transactions on Parallel and Distributed Systems},
  vol.~31, no.~2, pp. 279--293, 2019.

\bibitem[Sbîrlea et~al.(2014)Sbîrlea, Budimlić, and Sarkar]{s2014Bounded}
J.~Sbîrlea, Z.~Budimlić, and V.~Sarkar, ``Bounded memory scheduling of
  dynamic task graphs,'' in \emph{International Conference on Parallel
  Architecture \& Compilation Techniques}, 2014.

\bibitem[Sergent et~al.(2016)Sergent, Goudin, Thibault, and
  Aumage]{Sergent2016Controlling}
M.~Sergent, D.~Goudin, S.~Thibault, and O.~Aumage, ``Controlling the memory
  subscription of distributed applications with a task-based runtime system,''
  in \emph{IEEE International Parallel \& Distributed Processing Symposium
  Workshops}, 2016.

\bibitem[Tsai et~al.(2013)Tsai, Fang, and Chou]{tsai2013optimized}
J.-T. Tsai, J.-C. Fang, and J.-H. Chou, ``Optimized task scheduling and
  resource allocation on cloud computing environment using improved
  differential evolution algorithm,'' \emph{Computers \& Operations Research},
  vol.~40, no.~12, pp. 3045--3055, 2013.

\bibitem[Ergu et~al.(2013)Ergu, Kou, Peng, Shi, and Shi]{ergu2013analytic}
D.~Ergu, G.~Kou, Y.~Peng, Y.~Shi, and Y.~Shi, ``The analytic hierarchy process:
  task scheduling and resource allocation in cloud computing environment,''
  \emph{The Journal of Supercomputing}, vol.~64, no.~3, pp. 835--848, 2013.

\bibitem[Praveenchandar and Tamilarasi(2021)]{praveenchandar2021dynamic}
J.~Praveenchandar and A.~Tamilarasi, ``Dynamic resource allocation with
  optimized task scheduling and improved power management in cloud computing,''
  \emph{Journal of Ambient Intelligence and Humanized Computing}, vol.~12,
  no.~3, pp. 4147--4159, 2021.

\bibitem[Brucker and Schlie(1990)]{brucker1990job}
P.~Brucker and R.~Schlie, ``Job-shop scheduling with multi-purpose machines,''
  \emph{Computing}, vol.~45, no.~4, pp. 369--375, 1990.

\bibitem[{\"O}zg{\"u}ven et~al.(2010){\"O}zg{\"u}ven, {\"O}zbak{\i}r, and
  Yavuz]{ozguven2010mathematical}
C.~{\"O}zg{\"u}ven, L.~{\"O}zbak{\i}r, and Y.~Yavuz, ``Mathematical models for
  job-shop scheduling problems with routing and process plan flexibility,''
  \emph{Applied Mathematical Modelling}, vol.~34, no.~6, pp. 1539--1548, 2010.

\bibitem[Roshanaei et~al.(2013)Roshanaei, Azab, and
  ElMaraghy]{roshanaei2013mathematical}
V.~Roshanaei, A.~Azab, and H.~ElMaraghy, ``Mathematical modelling and a
  meta-heuristic for flexible job shop scheduling,'' \emph{International
  Journal of Production Research}, vol.~51, no.~20, pp. 6247--6274, 2013.

\bibitem[Birgin et~al.(2014)Birgin, Feofiloff, Fernandes, De~Melo, Oshiro, and
  Ronconi]{birgin2014milp}
E.~G. Birgin, P.~Feofiloff, C.~G. Fernandes, E.~L. De~Melo, M.~T. Oshiro, and
  D.~P. Ronconi, ``A milp model for an extended version of the flexible job
  shop problem,'' \emph{Optimization Letters}, vol.~8, no.~4, pp. 1417--1431,
  2014.

\bibitem[Hansmann et~al.(2014)Hansmann, Rieger, and
  Zimmermann]{hansmann2014flexible}
R.~S. Hansmann, T.~Rieger, and U.~T. Zimmermann, ``Flexible job shop scheduling
  with blockages,'' \emph{Mathematical Methods of Operations Research},
  vol.~79, no.~2, pp. 135--161, 2014.

\bibitem[Zhang and Zhou(2017)]{zhang2017dynamic}
P.~Zhang and M.~Zhou, ``Dynamic cloud task scheduling based on a two-stage
  strategy,'' \emph{IEEE Transactions on Automation Science and Engineering},
  vol.~15, no.~2, pp. 772--783, 2017.

\bibitem[Xia et~al.(2019)Xia, Quek, Zhang, Jin, and Zhu]{xia2019programmable}
W.~Xia, T.~Q. Quek, J.~Zhang, S.~Jin, and H.~Zhu, ``Programmable hierarchical
  c-ran: From task scheduling to resource allocation,'' \emph{IEEE Transactions
  on Wireless Communications}, vol.~18, no.~3, pp. 2003--2016, 2019.

\bibitem[Fu et~al.(2020)Fu, Tang, Yang, and Liu]{fu2020optimal}
Z.~Fu, Z.~Tang, L.~Yang, and C.~Liu, ``An optimal locality-aware task
  scheduling algorithm based on bipartite graph modelling for spark
  applications,'' \emph{IEEE Transactions on Parallel and Distributed Systems},
  vol.~31, no.~10, pp. 2406--2420, 2020.

\bibitem[Yuan et~al.(2018)Yuan, Bi, and Zhou]{yuan2018spatial}
H.~Yuan, J.~Bi, and M.~Zhou, ``Spatial task scheduling for cost minimization in
  distributed green cloud data centers,'' \emph{IEEE Transactions on Automation
  Science and Engineering}, vol.~16, no.~2, pp. 729--740, 2018.

\bibitem[Hu et~al.(2020)Hu, Li, and He]{hu2020reformed}
Y.~Hu, J.~Li, and L.~He, ``A reformed task scheduling algorithm for
  heterogeneous distributed systems with energy consumption constraints,''
  \emph{Neural Computing and Applications}, vol.~32, no.~10, pp. 5681--5693,
  2020.

\bibitem[Zhao et~al.(2021)Zhao, Li, Bao, Luo, He, and Liu]{zhao2021fairness}
M.~Zhao, W.~Li, L.~Bao, J.~Luo, Z.~He, and D.~Liu, ``Fairness-aware task
  scheduling and resource allocation in uav-enabled mobile edge computing
  networks,'' \emph{IEEE Transactions on Green Communications and Networking},
  vol.~5, no.~4, pp. 2174--2187, 2021.

\bibitem[Zhuge et~al.(2012)Zhuge, Guo, Hu, Tseng, Xue, and
  Sha]{zhuge2012minimizing}
Q.~Zhuge, Y.~Guo, J.~Hu, W.-C. Tseng, C.~J. Xue, and E.~H.-M. Sha, ``Minimizing
  access cost for multiple types of memory units in embedded systems through
  data allocation and scheduling,'' \emph{IEEE Transactions on Signal
  Processing}, vol.~60, no.~6, pp. 3253--3263, 2012.

\bibitem[Zuo et~al.(2013)Zuo, Zhang, and Tan]{zuo2013self}
X.~Zuo, G.~Zhang, and W.~Tan, ``Self-adaptive learning pso-based deadline
  constrained task scheduling for hybrid iaas cloud,'' \emph{IEEE Transactions
  on Automation Science and Engineering}, vol.~11, no.~2, pp. 564--573, 2013.

\bibitem[Shao et~al.(2005)Shao, Zhuge, Xue, and Sha]{shao2005efficient}
Z.~Shao, Q.~Zhuge, C.~Xue, and E.-M. Sha, ``Efficient assignment and scheduling
  for heterogeneous dsp systems,'' \emph{IEEE Transactions on Parallel and
  Distributed Systems}, vol.~16, no.~6, pp. 516--525, 2005.

\bibitem[Ding et~al.(2019)Ding, L{\"u}, Li, Shen, Xu, and Glover]{ding2019two}
J.~Ding, Z.~L{\"u}, C.-M. Li, L.~Shen, L.~Xu, and F.~Glover, ``A two-individual
  based evolutionary algorithm for the flexible job shop scheduling problem,''
  in \emph{Proceedings of the AAAI Conference on Artificial Intelligence},
  vol.~33, 2019, pp. 2262--2271.

\bibitem[Gonz\'alez et~al.(2015)Gonz\'alez, Vela, and Varela]{Gonz2015Scatter}
M.~A. Gonz\'alez, C.~R. Vela, and R.~Varela, ``Scatter search with path
  relinking for the flexible job shop scheduling problem,'' \emph{European
  Journal of Operational Research}, vol. 245, no.~1, pp. 35--45, 2015.

\bibitem[Mastrolilli and Gambardella(2000)]{mastrolilli2000effective}
M.~Mastrolilli and L.~M. Gambardella, ``Effective neighbourhood functions for
  the flexible job shop problem,'' \emph{Journal of Scheduling}, vol.~3, no.~1,
  pp. 3--20, 2000.

\bibitem[Gupta and Garg(2017)]{Gupta2017Load}
A.~Gupta and R.~Garg, ``Load balancing based task scheduling with {ACO} in
  cloud computing,'' in \emph{2017 International Conference on Computer and
  Applications (ICCA)}, 2017.

\end{thebibliography}






\end{document}